\documentclass[
superscriptaddress,
 amsmath,amssymb,
 aps,
 prb,
 twocolumn
]{revtex4-2}

\usepackage{float}
\usepackage{graphicx}
\usepackage{braket}
\usepackage{tabularx,ragged2e,booktabs,caption}
\usepackage{amsmath}
\usepackage[scr=boondox]{mathalfa}
\usepackage{inputenc}
\usepackage{stix}
\usepackage[dvipsnames]{xcolor}
\usepackage[colorlinks,linkcolor=blue,citecolor=blue,urlcolor=blue]{hyperref}
              
\renewcommand{\vec}[1]{\mathbf{#1}}

\begin{document}

\preprint{APS/123-QED}

\title{Dynamical signatures of symmetry protected topology following symmetry breaking}

\author{Jacob A. Marks}
\email{jamarks@stanford.edu}
 \affiliation{Physics Department, Stanford University, Stanford, CA 94035, USA}
 \affiliation{Stanford Institute for Materials and Energy Sciences (SIMES),
	SLAC National Accelerator Laboratory, Menlo Park, CA 94025, USA}
\author{Michael Sch\"uler}%
 \email{schuelem@stanford.edu}
 \affiliation{Stanford Institute for Materials and Energy Sciences (SIMES),
SLAC National Accelerator Laboratory, Menlo Park, CA 94025, USA}%

\author{Thomas P. Devereaux}
\affiliation{Stanford Institute for Materials and Energy Sciences (SIMES),
SLAC National Accelerator Laboratory, Menlo Park, CA 94025, USA}
\affiliation{Department of Materials Science and Engineering, Stanford University, Stanford, CA 94035, USA}

\date{\today}

\begin{abstract}
We investigate topological signatures in the short-time non-equilibrium dynamics of symmetry protected topological (SPT) systems starting from initial states which break a protecting symmetry. Naïvely, one might expect that topology loses meaning when a protecting symmetry is broken. Defying this intuition, we illustrate, in an interacting Su-Schrieffer-Heeger (SSH) model, how this combination of symmetry breaking and quench dynamics can give rise to both single-particle and many-body signatures of topology. From the dynamics of the symmetry broken state, we find that we are able to dynamically probe the equilibrium topological phase diagram of a symmetry respecting projection of the post-quench Hamiltonian. In the ensemble dynamics, we demonstrate how spontaneous symmetry breaking (SSB) of a protecting symmetry can result in a quantized many-body topological `invariant' which is not pinned under unitary time evolution. We dub this `dynamical many-body topology' (DMBT). We show numerically that both the pure state and ensemble signatures are remarkably robust, and argue that these non-equilibrium signatures should be quite generic in SPT systems, regardless of protecting symmetries or spatial dimension.
\end{abstract}

\maketitle
\section{Introduction}
Out-of-equilibrium dynamics represents a burgeoning frontier of physics research~\cite{Abanin_2019, Swingle_2016, Nandkishore_2015, Khemani_2016, Alba_2017, Choi_2020}. Certain non-equilibrium phenomena can be seen as dynamical counterparts to well-understood equilibrium phenomena~\cite{Heyl_2018}. Others have no equilibrium analog, presenting us with novel physics and providing new avenues for the control and manipulation of quantum phases of matter. Recent advances in atomic physics provide unprecedented ability to investigate and test these phenomena in the lab.

Topology, as a novel framework for understanding entanglement and characterizing order in quantum states, has garnered tremendous interest in the physics community~\cite{hasan_colloquium:_2010, qi_rmp_2011}. The pioneering work of Kitaev and others has paved the way for the systematic classification of topological phases in the equilibrium context, of both the non-interacting and the interacting variety~\cite{Kitaev_2009, Wang_2014, Isobe_2015, classif_3d}.

Recently, the synthesis of the these two areas - the study of topology out of equilibrium - has gained traction due to the potential for non-equilibrium engineering of topological properties. Great effort has been undertaken to devise comprehensive topological classification schemes, to identify signatures of topology of out equilibrium, and to characterize the relations between these signatures~\cite{mcginley_classif_2019,  mcginley_interacting_spt_noneq_2019, McGinley_1d_2018, Schuler_2017, Sun_2018, PhysRevLett.124.160402, Tarnowski_2019, Tsomokos_2009, Kells_2014, PhysRevLett.113.045303,Pastori_2020, PhysRevB.100.041101}.

In this work, we add a new wrinkle to this story by identifying and thoroughly characterizing post-quench signatures of topology in symmetry protected topological (SPT) systems starting from a symmetry broken initial state. While recent works have investigated certain aspects of the interplay between topology and symmetry breaking, to our knowledge the primary scenario discussed in this paper is previously undocumented~\cite{cuadra_prb_2019, cuadra_natcomm_2019, McGinley_1d_2018, phase_vortices}. We attribute this to the prior intuition that in SPT systems, concepts of topology lose meaning when the state of the system breaks a protecting symmetry. Our work connects to recent experimental efforts in realizing topological phases with ultracold atoms in optical lattices~\cite{aidelsburger_experimental_2011, struck_tunable_2012, goldman_topological_2016, dauphin_loading_2017, flaschner_experimental_2016}, and potential verification schemes could benefit from the body of recent progress toward measuring topological properties~\cite{Atala_2013, Wang_2016, Zheng_2020, Pe_a_Ardila_2018, Aidelsburger_2014, Elben_2020, dehghani2020extraction}.

Much recent work has been devoted to understanding topology out of equilibrium in non-interacting systems. Scenarios involving dynamical topological phase transitions (in the thermodynamic limit) and equilibration of topological properties (in the presence of thermalizing interaction) have been thoroughly investigated~\cite{PhysRevB.93.085416, kruckenhauser_dynamical_2018}. Additionally, quenches from symmetry broken states in (intrinsically topological) Chern insulators have been shown to enable dynamical probing of the equilibrium topological phase diagram~\cite{phase_vortices}. All of these works focused solely on post-quench signatures of topology on the single-particle level. This is because many-body topological invariants of symmetry respecting initial states are pinned under unitary time evolution. All of these signatures are also discrete-valued, fitting the traditional picture of topology.

The non-equilibrium interplay between SPT and breaking of protecting symmetries was assumed to work against topology, if not destroy it entirely. This is best exemplified by the scenario in~\cite{McGinley_1d_2018}, wherein states which initially obey the protecting symmetries dynamically break anti-unitary symmetries such as time-reversal symmetry, and thereby reduce the topological classification.

In this work, focusing on a variant of the Su-Schrieffer-Heeger (SSH) model~\cite{su_solitons_1979, heeger_solitons_1988}, we consider quench scenarios in which both the pre-quench (ground state) Hamiltonian $\hat{H}_0$ and the post-quench Hamiltonian $\hat{H}_1$ obey a set of protecting symmetries, but the initial state breaks one of the protecting symmetries. On the single-particle level, this gives rise to signatures of the ground state topology of $\hat{H}_1$ (at critical times) in the fractionalized but mobile Zak phase, and correspondingly in the (many-body) Resta polarization of the state. These signatures are even robust to the presence of small terms in $\hat{H}_1$ which explicitly break the protecting symmetries. Moreover, these dynamical topological signatures are not due to thermalization or equilibration, but instead occur in the short-time dynamics.

In the case of spontaneous symmetry breaking (SSB), we show how symmetry and dynamics conspire to produce dynamical many-body topological (DMBT) signatures, circumventing the many-body topological constraint. These many-body signatures inherit their mobility from the dynamics of the symmetry broken states, and their quantization from the symmetry relating the two ground states.

\begin{figure}[t!]
\includegraphics[width=0.48\textwidth]{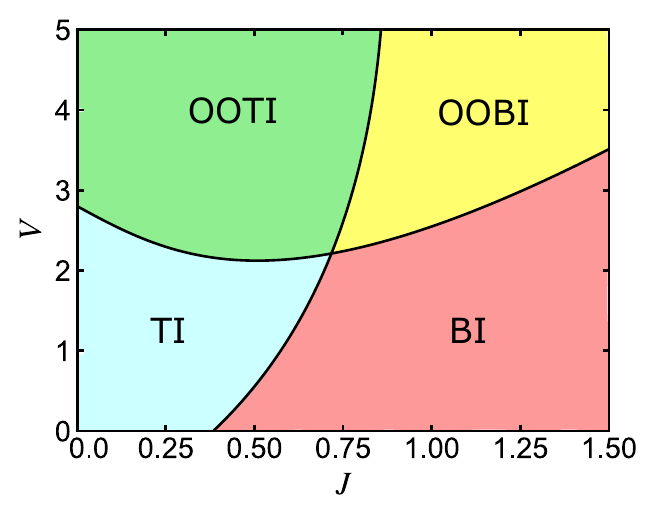} 
\caption{\label{fig:equilib_phasediagram}Sketch of a slice of the equilibrium phase diagram for interacting SSH model~(\ref{eq:model}), where $d = 0.4$ and $\tau = 1.0$, as detailed in~\cite{marks2020correlationassisted}. The phase diagram hosts four phases; the topological insulating (TI) and band insulating (BI) phases, and their order-obstructed analogs, OOTI and OOBI, in which the topology is obstructed by a charge density wave. The phase diagram is qualitatively unchanged for different values of $d$ and $\tau$, so long as $\tau \neq 0$. Quantized charge pumping is possible only in the TI and OOTI phases.}
\end{figure}

The paper is organized as follows. In Sec.~\ref{sec:model} we introduce the toy model and quench protocol considered in the paper. We also review three related but distinct topological invariants spanning single-particle and many-body notions of topology, namely the Zak phase, Resta polarization, and an invariant extracted from the single-particle density matrix (SPDM). We also discuss the differences between pure state and ensemble topology. In Sec.~\ref{sec:pure_state_sigs} we systematically investigate the signatures of topology in the pure state post-quench dynamics following symmetry breaking. We first introduce (Sec.~\ref{subsec:ps_setup}) the requisite single-particle machinery and define relevant quantities, including a dynamical contribution to the Zak phase. We develop intuition by analytically treating various limiting cases (Sec.~\ref{subsec:ps_TI}-\ref{subsec:ps_dlt1}), finding that
at certain critical times post-quench, the Zak phase approximately coincides with the equilibrium Zak phase of the post-quench Hamiltonian. We then refine this intuition (Sec.~\ref{subsec:ps_phasediagram}) and find that in more general cases, cusps in the Zak phase (and Resta polarization) reveal the ground state topology of the post-quench Hamiltonian. We use this observation to develop protocols for recovering equilibrium topology from pure state quench dynamics. In Sec.~\ref{sec:zte_sigs}, we study topological signatures in the ensemble quench dynamics. First (Sec.~\ref{subsec:zte_setup}) we characterize the constraints imposed on the Zak phase and Resta polarization of the ensemble in the post-quench dynamics. We show that while the Zak phase is pinned (without thermalizing interaction), the Resta polarization becomes mobile while retaining quantization. We then look at the short-time dynamics (Sec.~\ref{subsec:zte_shorttime}) to argue for the presence of potential topological behavior in the post-quench dynamics. Finally, (Sec.~\ref{subsec:zte_stability}) we define a measure of dynamical topological stability, and verify numerically that the distance of the post-quench Hamiltonian from the equilibrium topological phase boundary determines this stability, justifying the interpretation of the ensemble Resta polarization as a non-equilibrium topological signature.

\section{Model and Methods}
\label{sec:model}
In this work, we study an interacting topological insulator described by the microscopic Hamiltonian:
\begin{align} \label{eq:model}
\hat{H}(J, d, \tau, V)=&- \sum_j \left( J \hat{b}^{\dag}_j \hat{a}_j +  \frac{d + \tau}{2} \hat{b}^{\dag}_j \hat{a}_{j+1} +  \frac{d - \tau}{2} \hat{a}^{\dag}_j \hat{b}_{j+1}\right) \nonumber \\ 
& + \mathrm{h.\,c.} + \quad V \sum_j \left(\hat{n}^a_j \hat{n}^b_{j + 1} + \hat{n}^b_j \hat{n}^a_{j + 1} \right),
\end{align}

as detailed in~\cite{marks2020correlationassisted}. This model~(\ref{eq:model}) is a symmetry protected topological (SPT) model, meaning that in equilibrium, its topological character (here a $\mathbb{Z}_2$ topological invariant) is \textit{protected} by a set of discrete symmetries, in this case chiral, time-reversal, and sublattice symmetry. SPT systems can exhibit only short-range entanglement and, unlike systems with intrinsically topological order, are only robust to perturbations which respect the protecting symmetries.

This model~(\ref{eq:model}) represents by no means the most general SPT Hamiltonian possible. For our purposes however, it strikes an ideal balance between generality and tractability. First, the SSH model without interaction (setting $V = 0$, $d= \tau$ above) is the paradigmatic one-dimensional Chern insulator. Second, the full model~(\ref{eq:model}) supports charge pumping. And third, the many-body character can be taken fully into account by both Exact Diagonalization (ED) and Density Matrix Renormalization Group (DMRG) techniques. 

Throughout this manuscript, we use the Greek $\ket{\psi}$ to signify many-body states, and script Latin letters like $\ket{\mathscr{u}}$ to denote single-particle states. Moreover, we follow the convention of using $\hat{\cdot}$ only for many-body operators, as in~(\ref{eq:model}), and for unit vectors, where the context is obvious. We also use boldface to denote vectors.

Model~(\ref{eq:model}) represents interacting spinless fermions on a bipartite chain, with $L$ sites, and orbitals \textit{a} and \textit{b} on each site. $J$ represents the intra-cell hopping, while $(d\pm\tau)/2$ represent first and third neighbor inter-cell hopping. We only consider the case of half-filling. The final term, $\hat{V}$, is a density-density interaction term which preserves the aforementioned symmetries.

The phase diagram of the model~(\ref{eq:model}), as depicted in Fig.~\ref{fig:equilib_phasediagram}, supports four distinct phases: a band insulating (BI) phase, a topological insulating (TI) phase, and order obstructed analogs for both (OOBI) and (OOTI), as we explain below.

In the absence of interaction ($V = 0$), the model reduces to an exactly solvable two-band model, wherein all state information is captured by the SPDM $\rho$ (as differentiated from the many-body density matrix $\hat{\rho}$). Alternatively, the pseudospin vector, represented in momentum-space, contains the same information. In momentum-space, the Hamiltonian can be written as 
\begin{align}
    H(V = 0) = \sum_k \vec{c}^{\dag}\big[\vec{h}_k \cdot \vec{\sigma} \big] \vec{c},
    \label{eq:sp_ham}
\end{align}
where $\vec{c} = (c_{k, a}, c_{k, b})$ is the spinor of sublattice fermionic annihilation operators and $\vec{\sigma}$ is the vector of Pauli matrices. The single particle Hamiltonian, $\vec{h}_k$, is defined according to $\vec{h}_k = (h_k^x, h_k^y, h_k^z)$, with $h_k^x = J + d \cos{(k)}$ and $h_k^y = \tau \sin{(k)}$. $h_k^z = 0$ reflects the chiral symmetry of the model. From this representation, we can directly read out the topology of the state: for $|J| < |d|$, the system is in TI, whereas $|J| > |d|$ corresponds to BI. The model is solved in momentum-space, with single-particle eigenstates

\begin{align}
    \ket{\mathscr{u}_k} = \frac{1}{\sqrt{2}}\begin{pmatrix} \pm e^{-i \phi_k} \\ 1 \end{pmatrix},
    \label{eq:ssh_gs}
\end{align}
where $\phi_k = \arg{(h_k^x + i h_k^y)}$.

The TI and BI phases are robust to small interactions (and perturbations) that respect the protecting symmetries, and extend into the moderate interaction regime. In fact, for weak interactions that do not spontaneously break chiral symmetry, the system can be described by a pseudospin vector of the same form as in~(\ref{eq:sp_ham}), with renormalized $d$ and $\tau$, ($\tilde{d}$, $\tilde{\tau}$ respectively). For large interaction an Ising-like charge density wave (CDW) order arises, with order parameter $\hat{S}^z_j = \hat{n}^b_j - \hat{n}^a_j$, with a pair of degenerate ground states, denoted $\ket{\psi^{A}}$ and $\ket{\psi^{B}}$, which have occupation predominantly on sublattice \textit{a} and \textit{b} respectively. The CDW order thus spontaneously breaks the protecting sublattice symmetry, and thereby obstructs the topology. This results in the two order obstructed phases (OOBI and OOTI), both of which are strongly-correlated phases. OOBI and OOTI are considered to be order obstructed in a symmetry protected topological sense; explicitly suppressing the CDW order in each phase recovers the corresponding SPT phase. Surprisingly, as documented in~\cite{marks2020correlationassisted}, certain aspects of the ground state topology survive in the presence of this obstructing CDW order~\footnote{In particular, quantized charge pumping reveals the underlying topology via adiabatic cyclic evolution. In fact, the obstructing order can even \textit{drive} charge pumping.}.

To comprehensively characterize the topology of the model across the phase diagram, we differentiate between state and ensemble topology. In particular, for the pure state $\ket{\psi^{A(B)}}$, where CDW order obstructs the topology, we have the density matrix

\begin{align}
    \label{eq:rho_ps}
    \hat{\rho}^{A(B)} = \ket{\psi^{A(B)}}\bra{\psi^{A(B)}}.
\end{align}

We will use the superscript `PS' for pure state when either state will suffice. In contrast, CDW order is explicitly suppressed in the zero-temperature thermal ensemble (ZTE), which is given by

\begin{align}
    \label{eq:rho_zte}
    \hat{\rho}^{ZTE} = \frac{1}{2}\big( \ket{\psi^A}\bra{\psi^A} + \ket{\psi^B}\bra{\psi^B}\big).
\end{align}

Practically, both the PS and ZTE can be realized. For the ZTE, we note that CDW ordering only occurs at $T = 0$~\footnote{According to the Mermin-Wagner theorem, this is the case for all one-dimensional quantum systems}, so the ZTE describes the system at arbitrarily low temperatures. Thermal ensembles at small $T$ will obey the same symmetries of the model and will exhibit the same qualitative behavior as the ZTE, albeit with some of the sharp topological features slightly smoothed out. The pure state can be obtained by applying a small pinning field $\hat{\Delta} = \Delta \sum_j \hat{S}^z_j$ which explicitly breaks the sublattice symmetry. The pure state and ensemble SPDMs can be extracted from the respective many-body density matrices~(\ref{eq:rho_ps}) and ~(\ref{eq:rho_zte}).

We also distinguish between single-particle and many-body topology, which differ in their ability to treat interactions. Many definitions of single-particle and many-body topological invariants exist in the literature~\cite{tknn_1982, PhysRevLett.51.51, Shapourian_2017, Hung_2014}. Here we consider the Zak phase as our prototypical single-particle invariant, and the Resta polarization as our representative many-body topological invariant, both of which are valid for systems with periodic boundary conditions (PBC)~\cite{zak_berrys_1989, resta_quantum-mechanical_1998}. Alternative single-particle and many-body topological invariants can be chosen without modifying the main findings. The primary constraint is that in equilibrium, the `invariant' can generically only be quantized when the protecting symmetries are respected. For completeness, we also provide an example of an equilibrium single-particle topological invariant (in this case extracted from the SPDM) which fails to meet these requirements. For all three topological signatures, the density matrix (and SPDM), as expressed in~(\ref{eq:rho_ps}) and~(\ref{eq:rho_zte}) implicitly captures the information about symmetry breaking. 

The Zak phase, in the convention we will use, is a property of the state as opposed to the Hamiltonian. It can be computed according to 
\begin{align}
    \mathcal{Z} = \frac{i}{\pi}\oint dk \langle \mathscr{u}_k | \partial_k |\mathscr{u}_k\rangle,
    \label{eq:zak}
\end{align}

where $\ket{\mathscr{u}_k}$ denote natural orbitals (momentum-space eigenstates of the SPDM), which reduce to Bloch states in the non-interacting case. The Zak phase can be interpreted as a Berry phase, summing over the lower (occupied) band, with the path in momentum-space taken over one reciprocal lattice cell. For the single-particle solutions given by~(\ref{eq:ssh_gs}), the Zak phase simplifies to 
\begin{align}
    \mathcal{Z} = \frac{1}{2\pi}\oint{dk \partial_k \phi_k},
    \label{eq:ssh_zak}
\end{align}
which takes on quantized values of $\mathcal{Z} \in \{ \pm 1, 0\}$.

In contrast to the Zak phase, which is defined in terms of the SPDM, the Resta polarization is defined directly in terms of the many-body density matrix~\cite{resta_quantum-mechanical_1998}
\begin{align}
P = \frac{q a_0}{2 \pi} \mathrm{Im} \ln \mathrm{Tr}\left[\hat{\rho}\, e^{\frac{i 2 \pi}{L a_0} \hat{X}}\right] \quad (\mathrm{mod} \,q a_0) \ ,
\label{eq:restapol}
\end{align}
where $a_0$ is the lattice spacing, $L$ is the number of unit cells, and $\hat{X} = \sum_i x_i \hat{n}_i$ is the many-body center of mass operator. The Resta polarization can be viewed as a single-point Berry phase, and like the Zak phase, it is a property of the state~\cite{pol_berry_phase, Resta_1998_qm_pos_extended}. 

For comparison with the Zak phase and Resta polarization, we introduce the additional topological invariant $\nu$ derived from the SPDM. In particular, the SPDM in momentum space, $\rho_k$, can be decomposed as $\rho_k = \frac{1}{2} \big[ 1 - \vec{S}_k \cdot \vec{\sigma} \big]$, where $\vec{\sigma}$ is the vector of Pauli matrices, and $\vec{S}_k$ is the pseudospin vector, which satisfies

\begin{align}
\label{eq:sz_realspace}
\langle \hat{S}^z\rangle = \frac{1}{L}\sum_k \langle \hat{S}^z_k\rangle.   
\end{align}

We can then extract the topological invariant $\nu$ according to

\begin{align}
    \label{eq:nu}
    (-1)^{\nu} = \mathrm{sgn}(S^x_{\Gamma} \, S^x_{\pi}),
\end{align}

where subscripts $\Gamma$ and $\pi$ refer to the corresponding points in the Brillouin zone. Even in equilibrium, while $\nu$ only contains topological information if the system Hamiltonian is symmetry respecting, by definition it is always quantized, $\nu \in \{ 0, 1 \}$. This crucial fact already prohibits $\nu$ from exhibiting short-time non-equilibrium signatures like those we find in $\mathcal{Z}$ and $P$.

In equilibrium, for non-interacting systems all three topological signatures align. Namely, $P = \frac{qa_0}{2}\nu = \frac{qa_0}{2} \mathcal{Z}$, with $\mathcal{Z} = 1$ for $|J| < |d|$ (TI phase) and $\mathcal{Z} = 0$ for $|J| > |d|$ (BI phase).
Note that the ground state is unique, so no distinction is made between pure state and ZTE. In the presence of strong ordering, where the ground states break sublattice symmetry, the ensemble values for all three quantities reveal the buried topology of the phase. However, the pure state values disagree. The Zak phase vanishes in the limit of large interaction, $\mathcal{Z}^{PS} \rightarrow 0$, for both OOBI and OOTI, and the SPDM invariant $\nu^{PS}$ generically does not reveal topology. The Resta polarization fractionalizes, $\lim_{V\rightarrow\infty}P^{PS} = \pm \frac{qa_0}{4}$, reflecting the two-fold ground state degeneracy. Importantly, because $P$ is many-body in nature, adiabatic charge pumping in the strongly interacting regime still reveals the buried topology.

We emphasize that each of these topological `invariants' is typically defined in an equilibrium context, and that strictly speaking, topology is ill-defined out of equilibrium. However, certain non-equilibrium scenarios have been documented under which these signatures still exhibit topological behavior. In the presence of a small thermalizing interaction, for instance, as described in~\cite{kruckenhauser_dynamical_2018}, $\mathcal{Z}$ and $\nu$ can dynamically adjust from capturing the topology of a pre-quench Hamiltonian to that of a post-quench Hamiltonian. This scenario, as well as the contrasting case of no thermalizing interaction, is included in Table~\ref{tab:quench_signatures}. The last three rows in the table summarize the new scenarios considered in this paper, which involve symmetry breaking. In this context, discrete and continuous refer to the possible values that can be taken by the signature, and pinned (as opposed to mobile) implies that the quantity retains its pre-quench value throughout the quench.

In addition to the base model~(\ref{eq:model}) we also consider the additional sublattice imbalance term $\hat{\Delta}$ which induces Ising-like order, allowing us to compare and contrast spontaneous and generic (non-spontaneous) symmetry breaking. When we include such a term, we explicitly parameterize the Hamiltonian as $\hat{H}(J, d, \tau, V, \Delta) = \hat{H}(J, d, \tau, V) + \hat{\Delta}$. This encompasses the full class of Hamiltonians we consider (pre-quench and post-quench). We investigate abrupt quenches wherein a system initially in equilibrium with respect to pre-quench (ground state) Hamiltonian $\hat{H}_0$ is propagated forward in time (from $t = 0$) according to post-quench Hamiltonian $\hat{H}_1$. 

For ease of notation (it will become apparent why), we use the shorthand $\hat{H}' = \hat{H}(J, d, \tau, V = 0, \Delta = 0)$ to refer to the symmetry respecting single-particle components of a generic many-body Hamiltonian under consideration. We also note that for an arbitrary Hamiltonian in this class, $\hat{H}(J, d, \tau, V, \Delta)$, the corresponding Hamiltonian with $\Delta = 0$, i.e. $\hat{H}(J, d, \tau, V, \Delta=0)$, is the closest topological representative~\footnote{We assume tacitly that $\hat{H}(J, d, \tau, V, \Delta=0)$ is not order-obstructed.}. The equilibrium (ground state) results detailed in Eqs.~(\ref{eq:sp_ham}),~(\ref{eq:ssh_gs}),~(\ref{eq:ssh_zak}) and the surrounding discussion will be referred to using the subscript (or superscript) `eq' to distinguish from quantities during the quench. Subscripts of `0' and `1' will refer to parameters of the pre~-~ and post~-~quench Hamiltonians, $\hat{H}_0$ and $\hat{H}_1$ respectively.

The accompanying numerical results shown in the main text are generated using ED ($L \leq 12$) for interacting systems and tight binding techniques ($L \geq 24$) for non-interacting systems, where we work directly in $k$~-~space, employing the continuum limit. Supporting short-time calculations on larger systems performed using DMRG techniques are included in the Appendix.

\section{Pure State Signatures}
\label{sec:pure_state_sigs}
The main result of this section is that, for systems with strong initial symmetry breaking, and relatively weak perturbations in quench interaction ($V_1$, $\Delta_1$ small), one can recover the equilibrium phase diagram of $\hat{H}_1'$ from signatures in the short-time post-quench dynamics of the pure state Zak phase and Resta polarization. In particular, out of equilibrium, $\mathcal{Z}^{PS}$ and $P^{PS}$ are both continuous-valued and mobile during the quench. We find that both signatures exhibit oscillatory behavior with \textit{cusps}. Moreover, the value of these signatures at said cusps reflects the equilibrium topology of $\hat{H}_1'$, the symmetry respecting portion of the post-quench Hamiltonian. These results are detailed in the third row of Table.~\ref{tab:quench_signatures}. Remarkably, to understand these results, we only need to take a single-particle perspective. 

In this section, we will focus on the case $\Delta_0 \rightarrow \infty$, wherein the ground state Hamiltonian explicitly breaks the protecting symmetries of the SSH model (equivalently, one of the two degenerate SSB states for $V_0 \rightarrow \infty$ with $\Delta_0 = 0$). In the Appendix, we address the slightly more involved case of finite $\Delta_0$ (or finite $V_0$ with $\Delta_0 = 0$) using the same techniques. So long as $\Delta_0$ (or $V_0$) is larger in scale than the other parameters in the pre-quench Hamiltonian, the same signatures are visible.

First, we analyze quenches with $\hat{H}_1 = \hat{H}_1'$ , (i.e. $\Delta_1 = V_1 = 0$), and show how the quench dynamics gives rise to signatures (in the cusps of $\mathcal{Z}^{PS}$ and $P^{PS}$) of the ground state topology of $\hat{H}_1$. Second, we introduce an explicit symmetry breaking term $\Delta_1$ into the post-quench Hamiltonian, and show that the topological signatures are robust (i.e. under the presence of small $\Delta_1$, the $\Delta_1= 0$ topological signatures persist. In other words, the non-equilibrium dynamics governed by $\hat{H}_1$ uncover the topology of $\hat{H}_1'$. Finally we validate these analytic results with extensive numerics, and find that, remarkably, these topological signatures enable dynamical recovery of the equilibrium topology of $\hat{H}_1'$ across all (considered) $J_1$, $d_1$, and $\tau_1$.

\subsection{Setup}
\label{subsec:ps_setup}
We begin with a simple symmetry broken state $\ket{\psi^{\Delta}}$, which is a ground state of pre-quench Hamiltonian $\hat{H}_0 = \hat{\Delta}$. On a single-particle level, this Hamiltonian can be written as $\vec{h}_0 = (0, 0, \Delta)$, and the initial state takes the form 

\begin{align}
    \label{eq:u_dlt}
    \ket{\mathscr{u}^{\Delta}_k} = \begin{pmatrix} 0 \\ 1 \end{pmatrix},
\end{align}

which is evidently independent of momentum. We reiterate that such a ground state can alternatively be produced (as one of two ground states) by a symmetry respecting Hamiltonian of the form $\hat{H}_0 = \hat{V}$. Due to the fact that $V_1 = 0$, the post-quench Hamiltonian, $\hat{H}_1$, can also be written in pseudospin language as $\vec{h}_{1}= (h^x_1, h^y_1, h^z_1)$, where the $k$~dependence has been omitted for readability.

As a function of time, the single-particle state can be described by 

\begin{align}
    \ket{\mathscr{u}_{k}(t)} = \begin{pmatrix} \mathscr{u}_{k, 0}(t) \\ \mathscr{u}_{k, 1}(t) \end{pmatrix} \sim \begin{pmatrix}
    i \sin{(h_1 t)}(\hat{h}_1^x - i \hat{h}_1^y) \\
    \cos{(h_1 t)} + i \sin{(h_1t)}\hat{h}_1^z  \end{pmatrix},
    \label{eq:sp_sb_quench}
\end{align}

according to~\cite{phase_vortices}, where $h = \sqrt{(h^x)^2 + (h^y)^2 + (h^z)^2}$ and $\hat{h}^{\alpha} = h^{\alpha}/h$. The momentum dependence of $h$ is left implicit.

Alternatively, the state can be parameterized on the Bloch sphere as

\begin{align}
    \ket{\mathscr{u}_{k}(t)} = \begin{pmatrix} \sin{(\theta_k/2)} \\ -\cos{(\theta_k/2)}e^{-i\phi_k} \end{pmatrix},
    \label{eq:sp_quench_bloch}
\end{align}

where 

\begin{align}
    \theta_k = \arccos{\Big( \cos^2{(h_1t)} + \sin^2{(h_1t)}[\hat{h}_{1, \perp}^2 - \hat{h}_{1, \parallel}^2] \Big)},
    \label{eg:theta}
\end{align}

and 

\begin{align}
    \phi_k = \arg{\Big[ \frac{\cos{(h_1 t)} + i \sin{(h_1t)}\hat{h}_{1, \perp}}{i \sin{(h_1t)}\hat{h}_{1, \parallel}} \Big]}.
    \label{eq:phi}
\end{align}

Here we have defined $h_{\perp} = h^z$ and $h_{\parallel} = h^x - i h^y$ to be the pseudospin components perpendicular and parallel to the $xy$ plane.

In this language, the instantaneous Zak phase at any time during the quench is expressed as 

\begin{align}
    \mathcal{Z}^{PS}(t) = \frac{1}{\pi}\oint{dk [\cos^2{(\theta_k/2)} \partial_k \phi_k]},
    \label{eq:zak_bloch}
\end{align}

for $t \neq 0$, which is generically not quantized. At $t= 0$ identically, $\mathcal{Z}^{PS} = 0$.

In equilibrium, the condition

\begin{align}
\langle \hat{S}^z_k \rangle_{eq} = 0 \, \, \forall k,
\label{eq:sz_k_cond}
\end{align}

is necessary (but not sufficient) for quantization of the Zak phase (and the Resta polarization). Guided by this equilibrium intuition, it is reasonable to ask if such a condition can be satisfied out of equilibrium. If such a condition is indeed satisfied, we can further ask what vestiges of topology can be found. To investigate these questions, we first note that the necessary condition~(\ref{eq:sz_k_cond}), applied during the quench, implies

\begin{align}
\rho_{k, 00}(t) =\rho_{k, 11}(t) = \frac{1}{2}.
\label{eq:rho_k_t_constraint}
\end{align}

Out of equilibrium, we are not guaranteed that~(\ref{eq:rho_k_t_constraint}) is satisfied for all $k$ simultaneously. As a corollary, we have no guarantees that the many-body state is an eigenstate of $\hat{H}_1$ for any $t$. Nevertheless, we proceed. Let $t^*(k^*)$ denote the first time at which Eq.~(\ref{eq:rho_k_t_constraint}) is satisfied for a given $k^*$. At this time, Eq.~(\ref{eq:sp_quench_bloch}) takes the form of~(\ref{eq:ssh_gs}) up to global phase factors. Moreover, we formally define 

\begin{align}
t^* = \frac{\int{t^*(k) dk }}{\int{dk}}
\label{eq:tstar}
\end{align}

to represent the time that minimizes the total deviation from Eq.~(\ref{eq:rho_k_t_constraint}) over all momenta $k$. Intuitively, this is the time at which the system is closest to obeying the (necessary) Zak phase quantization condition, and is thus closest to being an eigenstate of Eq.~(\ref{eq:ssh_gs}). Consequently, we hope to observe topological signatures at time $t^*$. We stress that, as we are dealing with pure states following symmetry breaking, the Resta polarization $P^{PS}$ is also unpinned. Therefore, if topological signatures are visible in the Zak phase at a given time, then similar signatures should propagate into the many-body dynamics, and should be visible in $P^{PS}$ at the same time. We note that we have defined $t^*$ as the $\textit{first}$ such time, because dephasing (and $k-$~dependence) build up over time, resulting in less accurate approximations with little if any additional insight.

We can gain further intuition for the presence of dynamical topological signatures by decomposing Eq.~(\ref{eq:phi}) into multiple constituent terms using properties of the $\arg{(\cdot)}$ function. Given the class of Hamiltonians under consideration, $h_1^x = h_{eq}^x$, and $h_1^y = h_{eq}^y$, allowing us to relate $\phi_k$ and $\phi^{eq}_k$ by

\begin{align}
\label{eq:phik_props}
\phi_k(t) &= \phi^{eq}_k +  \phi^{dyn}_k(t) - \frac{\pi}{2}\mod{2 \pi},
\end{align}

where we have defined  

\begin{align}
\label{eq:phik_dyn}
    \phi^{dyn}_k(t) = \arg{[\cot{(h_1t)} + i \hat{h}_1^z]},
\end{align}
which contains all of the time-dependence. The constant contribution vanishes under action of the gradient $\partial_k$. Using the double-angle identity, we can then rewrite Eq.~(\ref{eq:zak_bloch}) as 

\begin{align}
    \label{eq:zak_relation}
    \mathcal{Z}^{PS} &= \frac{1}{2\pi}\oint{dk \big( 1 + \cos{(\theta_k)}\big)\partial_k \phi_k } \\
    &= \mathcal{Z}^{eq} + \mathcal{Z}^{dyn},
\end{align}

where 

\begin{align}
    \label{eq:zak_dyn}
     \mathcal{Z}^{dyn} = \frac{1}{2 \pi}\oint{dk \Big[\partial_k\phi^{dyn}_k + \cos{(\theta_k)}\partial_k\big( \phi^{eq}_k + \phi^{dyn}_k\big) \Big]}
\end{align}

is the dynamical contribution to the Zak phase, and time-dependence is implicit. At arbitrary times, this dynamical contribution can be quite large, and looking at the instantaneous Zak phase will not provide much insight into the ground state topology. However as we will show below, at certain times this contribution becomes small and we can recover details of the ground state topology of $\hat{H}_1'$. Namely, the times at which this dynamical contribution becomes least important are precisely the times at which $\ket{\psi_k(t)}$ is closest to an eigenstate of $\hat{H}_1'$. We also stress that aside from the particular expressions for the state and the Zak phase, this formalism is completely general for non-interacting two-band models.

\subsection{Case I: Quenching deep into TI}
\label{subsec:ps_TI}
As a first scenario, we consider quenches with $\hat{H}_1 \in \mathrm{TI}$ far from the equilibrium topological phase boundary. This is a scenario in which we should see very pronounced topological signatures, if such signatures exist in the non-equilibrium dynamics at all. In particular, as $|J| < |d|$ defines the TI phase in equilibrium, we consider $J_1 \ll d_1$, and let $\epsilon = |J_1/d_1| \ll 1$ be a small parameter. We can then Taylor expand the general equations in Sec.~\ref{subsec:ps_setup} in the depth of the quench. Here, we need only Taylor expand up to first order in $\epsilon$. Importantly, we take $\tau_1 = d_1$, (representing a traditional SSH model with no third-neighbor hopping). In Sec.~\ref{subsec:ps_phasediagram}, we clarify which elements of this treatment generalize to $\tau_1 \neq d_1$. Additionally, we set $\Delta_1 = 0$ so that there are no explicit symmetry breaking terms, and also take $V_1 = 0$ so that $\hat{H}_1 = \hat{H}_1'$. We also note that, in the strict case $J_1 = 0$ or $d_1 = 0$, these arguments do not apply and the Zak phase is pinned to its initial value.

First, since $h_1^z = 0$, we note that $\mathscr{u}_{k, 1}(t) = \cos{(h_1 t)}$ and $|h_1| = |h_{1, \parallel}|$. Taken together, these give $\rho_{k, 00}(t) = \sin^2{(h_1t)}$, and $\rho_{k, 11}(t) = \cos^2{(h_1t)}$. Plugging these into Eq.~(\ref{eq:rho_k_t_constraint}), we arrive at $t = \frac{\pi}{4h_1}$. The single-particle Hamiltonian, $h_1 \approx d_1 [ 1 + \epsilon \cos{k}]$, is only weakly $k-$~dependent, (explicitly $t^*(k) \approx \frac{\pi}{4d_1}[1 - \epsilon \cos{(k)}]$), and at the point $t^* = \frac{\pi}{4d_1}$, we find that 
\begin{align}
\langle \hat{S}^z_k(t^*) \rangle_I = \frac{\pi}{2}\epsilon \cos(k),
\label{eq:sz_I}
\end{align}

so that for all $k$, the equilibrium Zak phase quantization condition approximately obtains. Moreover $\langle \hat{S}^z \rangle = 0$ to first order in $\epsilon$. These first order approximations are borne out with great accuracy in exact numerics. At $t = t^*$, the state is approximately an eigenstate of $\hat{H}_1$. We also see that $t^*$ is inversely related to $d_1$, which we have already assumed to be the largest scale in the system. Thus, we can make $d_1$ arbitrarily large, and $t^*$ correspondingly small.

What's more, the complex phase factor simplifies to $\phi^{dyn}_k(t) = \arg{(\cot{(h_1t)})}$, which is real for all $t$ such that $h_1t \neq \pi n$. This means that $\partial_k \phi^{dyn}_k(t) = 0 \, \, \forall t$, and consequently the dynamical contribution does \textit{not} enter the Zak phase. Additionally, we can simplify $\cos{\theta_k} = \cos{(2h_1t^*)}$ by expanding around $\frac{\pi}{2}$, leading us to 

\begin{align}
    \label{eq:zak_dyn_I}
    \mathcal{Z}^{dyn}_I(t^*) \approx -\frac{\epsilon}{4}\oint{dk \cos{(k)} \partial_k \phi^{eq}_k },
\end{align}

which contributes (apparently) linearly in $\epsilon$ to the instantaneous Zak phase. However, as we will see explicitly in the following scenario, the integral hides another factor of the small parameter (here $\epsilon$), making the scaling quadratic. Specifics of the scaling aside we see that, starting from a symmetry broken state and quenching deep into the topological phase, we can recover, at a particular time $t^*$, signatures of the equilibrium topology of the post-quench Hamiltonian. Moreover, the specific time $t^*$ is parametrically small in the system parameters.

\subsection{Case II: Quenching deep into BI}
\label{subsec:ps_BI}
Second, we consider another deep quench, but this time to the other side of the equilibrium topological phase boundary. In particular, we quench deep into the BI phase ($J_1 \gg d_1$), again with $\tau_1 = d_1$ and without any symmetry breaking terms ($\Delta_1 = 0$) or interaction ($V_1 = 0$) in $\hat{H}_1$. This time, we let $\delta = d_1/J_1 \ll 1$ be our small parameter, and expand to lowest order in this quantity. As in Sec.~\ref{subsec:ps_TI}, we see signatures of the equilibrium topology of the post-quench Hamiltonian.

The resulting equations look very similar to in case I. Namely, $h_1 \approx J_1(1 + \delta \cos{(k)})$, exhibiting the same weak $k-$~dependence, $t^*(k) \approx \frac{\pi}{4 J_1}[1 - \delta \cos{(k)}]$, and correspondingly $t^* = \frac{\pi}{4 J_1}$. At this time, the deviations from the quantization condition can be expressed as

\begin{align}
\label{eq:sz_II}
\langle \hat{S}^z_k(t^*) \rangle_{II} = \frac{\pi}{2}\delta \cos(k),
\end{align}

which has the same functional form as in case I. As in case I, $\langle \hat{S}^z(t^*)\rangle_{II} = 0$ to first order. The Zak phase also simplifies in an analogous manner, with 
\begin{align}
\label{eq:zak_dyn_II}
    \mathcal{Z}^{dyn}_{II}(t^*) & \approx -\frac{\delta}{4}\oint{dk \cos{(k)} \partial_k \phi^{eq}_k },
\end{align}

which is again apparently linear in the small quantity $\delta$. However, the apparent linear scaling with $\delta$ in~(\ref{eq:zak_dyn_II}) is deceptive, as it implicitly hides the $\delta$~-~dependence of the integral itself. In this case, we can further approximate the integral, giving 

\begin{align}
\label{eq:zak_dyn_II_2ndorder}
    \mathcal{Z}^{dyn}_{II}(t^*) & \approx -\frac{\pi \delta^2}{4},
\end{align}

the details for which can be found in the Appendix.

Additionally, we can utilize the fact that $h^x_1>0$ all along the closed path to upper bound the contribution according to 

\begin{align}
\label{eq:zak_dyn_II_bound}
    |\mathcal{Z}^{dyn}_{II}(t^*)| < \frac{\delta \pi^2}{2},
\end{align}

the derivation for which can be found in the Appendix.

Crucially, at least for deep quenches with only nearest-neighbor hopping, this means that we are able to distinguish the equilibrium topology of post-quench Hamiltonian $\hat{H}_1$.

\subsection{Case III: Adding in $\Delta_1$}
\label{subsec:ps_dlt1}

In Sec.~\ref{subsec:ps_TI} and~\ref{subsec:ps_BI}, we restricted our attention to cases in which $\hat{H}_1 = \hat{H}_1'$. Here, we demonstrate that the previous conclusions are robust to the addition of a sublattice imbalance term $\Delta_1 \neq 0$ that explicitly breaks chiral symmetry. For small enough $\Delta_1$, the topological signature at $t^*$ reflects the underlying topology of $\hat{H}_1'$. This means that, remarkably, even when the post-quench Hamiltonian (perturbatively) breaks a protecting symmetry, we are \textit{still} able to identify in the non-equilibrium dynamics some vestiges of topology. Now however, those topological signatures reflect the equilibrium topology of a different Hamiltonian - namely the symmetry respecting single-particle component of the post-quench Hamiltonian.

Here, we explicitly consider the case $J_1 \gg \Delta_1, d_1$ with $d_1 = \tau_1$, but analogous results for $J_1 \ll d_1$ can be derived by following the same procedure. As before, we let $\delta = |d_1/J_1|$ be a small quantity, and additionally we let $\chi = |\Delta_1/J_1|$ be a second small quantity. As $\Delta_1$ only appears (at lowest order) quadratically in $h_1$, we must expand to second order in $\chi$ to see the effect of the sublattice imbalance. Consequently, we also expand to second order in $\delta$, i.e. $\mathcal{O}(\delta^2, \chi^2, \delta \chi)$.

In this case, we have $h_1 \approx J_1 [ 1 + \delta \cos{(k)} + \frac{\chi^2 + \delta^2\sin^2{(k)}}{2}]$, with $\hat{h}_1^z \approx \chi [ 1 - \delta \cos{(k)}]$. The critical time from case II is modified to $t^*(k) = \frac{\pi}{4 J_1}[1 - \delta \cos{(k)} - \frac{\delta^2}{2}(1 + \cos^2{(k)})+ \frac{\chi^2}{2}(\frac{4}{\pi} - 1)]$, leading to $t^* = \frac{\pi}{4 J_1}[1 - \frac{3}{4}\delta^2 + \frac{\chi^2}{2}(\frac{4}{\pi} - 1)]$. Fortuitously, the $\chi$~-~dependence cancels out of $\langle \hat{S}^z_k(t^*) \rangle$, and~(\ref{eq:sz_II}) is unmodified (up to appropriate order), i.e.

\begin{align}
\label{eq:sz_III}
\langle \hat{S}^z_k(t^*) \rangle_{III} = \frac{\pi}{2}[\delta \cos(k) - \frac{1}{2}\delta^2(\cos^2{(k)} + \frac{1}{2})],
\end{align}

implying that at time $t^*$, the many-body states are approximately eigenstates of the closest symmetry respecting Hamiltonian $\hat{H}(J_1, d_1, \tau_1, 0, 0)$. The sublattice imbalance $\Delta_1$ has the effect of shifting the critical time $t^*$ slightly later, but does not strongly alter the degree to which the state at time $t^*$ approximates an eigenstate of $\hat{H}(J_1, d_1, \tau_1, 0, 0)$. 

Turning to the Zak phase, the correction with $\chi$ scales quartically, and can be reasonably ignored.  To see this, we first note that $\cos{(\theta_k)} = \cos{(2h_1t)} + 2 \hat{h}_{1,z}^2 \sin^2{(h_1t)}$. While $\hat{h}_{1,z} \neq 0$ results in a non-zero $\phi^{dyn}_k(t)$, its contribution to the path integral vanishes. The result is that Eq.~(\ref{eq:zak_dyn_II})~-~(\ref{eq:zak_dyn_II_bound}) apply directly to the case $\Delta_1 \neq 0$ (so long as $\Delta_1 \ll J_1$). Explicitly, 

\begin{align}
\label{eq:zak_dyn_III}
    \mathcal{Z}^{dyn}_{III}(t^*) & \approx -\frac{\delta}{4}\oint{dk \cos{(k)} \partial_k \phi^{eq}_k },
\end{align}

In other words, $\Delta_1$ has the effect of shifting $t^*$, but does not alter the instantaneous Zak phase at that time. 

To recap, starting from a symmetry breaking initial state, and evolving with a symmetry breaking post-quench Hamiltonian, we are still able to identify, in the non-equilibrium dynamics, signatures of the topology of an SPT Hamiltonian. And this SPT Hamiltonian is precisely the symmetry respecting single-particle component of the post-quench Hamiltonian. Of course, this does not hold for all possible symmetry breaking pre-quench and post-quench Hamiltonians. We are implicitly assuming that the symmetry breaking term in $\hat{H}_1$ is small - for $\Delta_1$ large enough, dynamical quantum phase transitions (DQPTs) in the order parameter dominate the post-quench behavior and wash away any topological signatures. Nevertheless, this is miraculous.

\subsection{Dynamically Probing Phase Diagram}
\label{subsec:ps_phasediagram}

\begin{figure*}[htb!]
\includegraphics[width=0.96\textwidth]{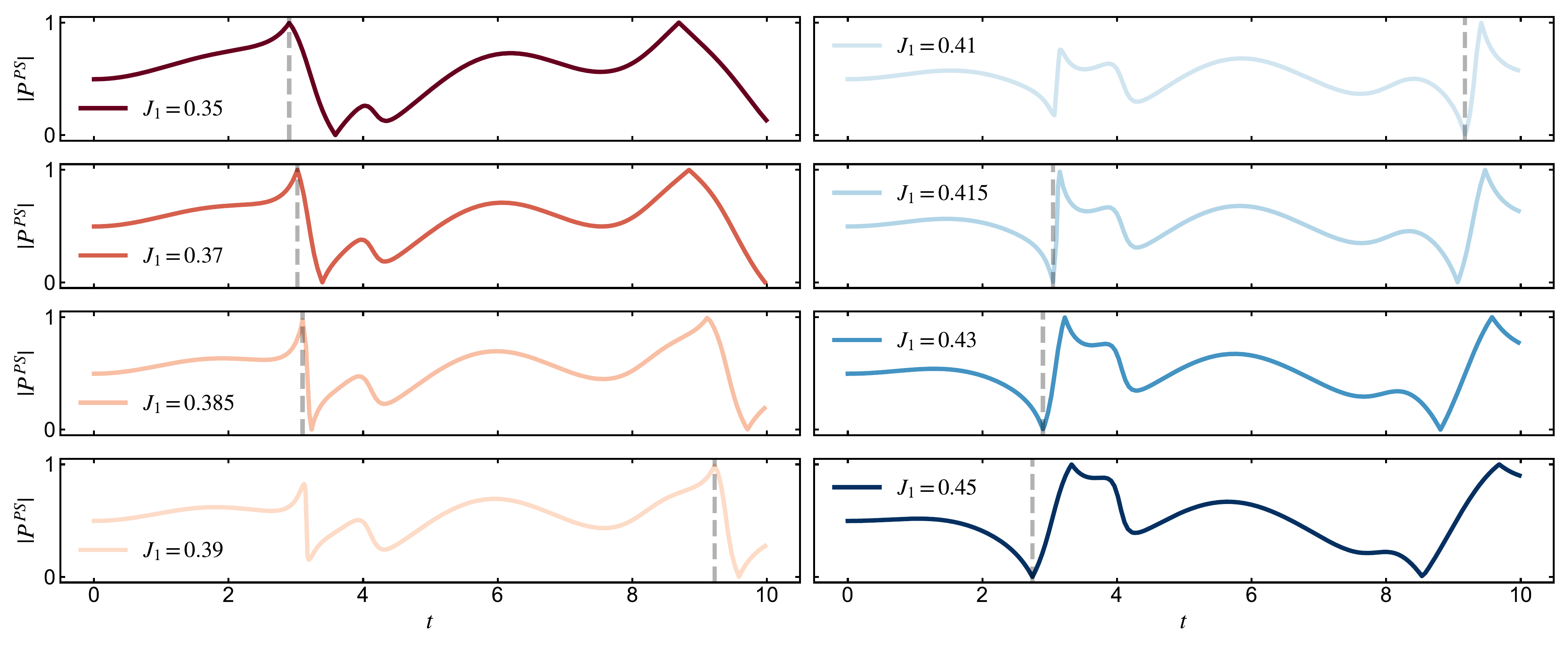} 
\caption{\label{fig:cusps}Illustration of the formation and deformation of topological and trivial cusps in the symmetry broken Resta polarization $|P^{PS}(t)|$ near equilibrium phase boundary, in units of $qa_0/2$. Here, $J_c = d_1 = 0.4$. The vertical dashed line on each axis represents the first genuine cusp, according to~(\ref{eq:crit3}). Data shown for systems of $L = 12$ unit cells with initial Hamiltonian given by $J_0 = 0.1$, $d_0 = 0.22$, $\tau_0 = 0.3$, $V_0 = 5.0$, and quench Hamiltonian given by $d_1 = 0.4$, $\tau_1 = 1.0$, $V_1 = 0.0$.}
\end{figure*}

\begin{figure}[htb!]
\includegraphics[width=0.48\textwidth]{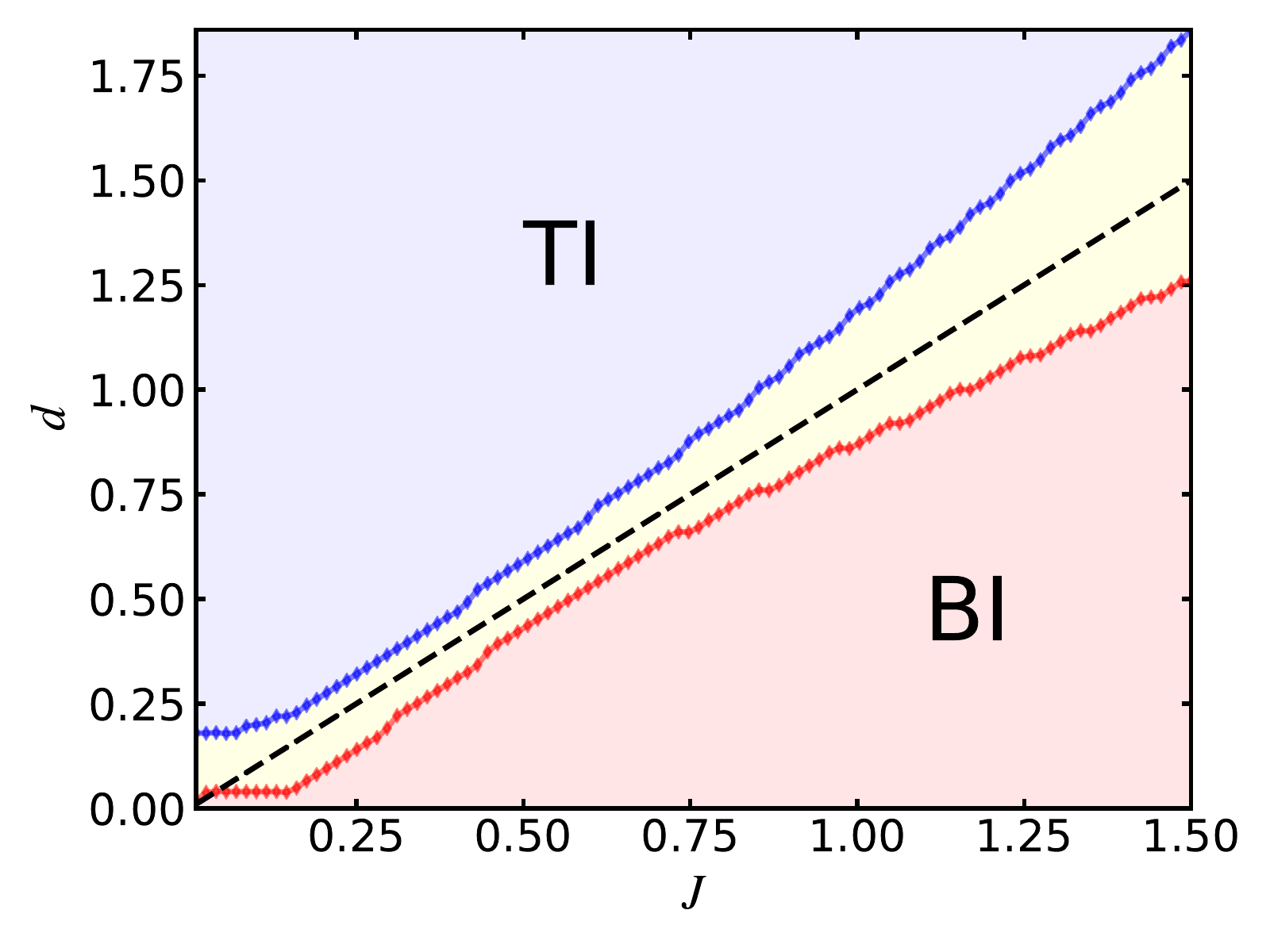} 
\caption{\label{fig:pol_cusp_phasediagram}Typical example of quench protocol for recovering the ground state topology of $\hat{H}_1'$ via cusps in the Resta polarization according to~(\ref{eq:crit3}). The black dashed line signifies the equilibrium phase boundary ($J = d$), and the yellow region encompassing it denotes the region of uncertainty given small system size ($L = 10$). Initial state was generated as the ground state of $\hat{H}_0$ with $\Delta_0 = 100 \tau_0$, and all other parameters set to zero. In the post-quench Hamiltonian $\hat{H}_1$,  with $\tau_1 = 1$ setting the scale, both sublattice imbalance $\Delta_1 =  0.1$ and interaction  $V_1 = 0.2$ are included. All quenches are stopped at $t_{max} = 20$. A threshold of $\eta = 0.1$ is used.}
\end{figure}

So far, the analytic results in Sec.~\ref{subsec:ps_TI}~-~\ref{subsec:ps_dlt1} have been limited to cases $J_1 \ll d_1$ or $J_1\gg d_1$, with $\tau_1 = d_1$ in all scenarios. With certain caveats, this physics also extends readily to less analytically tractable parameter regimes. Here, we explain which aspects of these phenomena generalize, and develop a heuristic procedure for dynamically characterizing the ground state topology of $\hat{H}_1'$ by performing quenches with $\hat{H}_1$. We simulate this procedure numerically using signatures in the Zak phase and the Resta polarization. For the Zak phase dynamics, we are able to employ tight binding methods, allowing us to reach large system sizes, resulting in almost perfect recovery of the equilibrium phase diagram of $\hat{H}_1'$. For the Resta polarization on the other hand, we are limited in our ED studies to $L \leq 12$ unit cells, resulting in a region of uncertainty surrounding the equilibrium topological phase boundary, as depicted in Fig.~\ref{fig:pol_cusp_phasediagram}.

For shallow quenches (away from $J \gg d$ and $d \gg J$), and for $\tau \neq d$, $\mathcal{Z}^{PS}$ and $P^{PS}$ still contain important topological information. However, as we demonstrate in the Appendix, for $\tau \neq d$ we cannot use $\langle \hat{S}^z(t)\rangle$ as a proxy to guide our intuition. Moreover, the analytic expressions implicitly utilized knowledge of the parameters of the post-quench Hamiltonian to determine $t^*$, the time at which to observe topological signatures. Here, we devise a more general criterion based on the Zak phase itself, informed by looking at the trajectory of the Zak phase over time during the quench. A similar criterion can be devised based on the Resta polarization. Importantly, the criteria based on the Zak phase and the Resta polarization do not assume any prior knowledge of system parameters.

Numerically, we find strong evidence that what generalizes (for Hamiltonians with third-neighbor hopping, close to phase boundary, and with interaction and sublattice imbalance) is the presence of `cusps' in $|\mathcal{Z}^{PS}(t)|$, as we will describe below. Similar cusps occur in $|P^{PS}|$, as illustrated in Fig.~\ref{fig:cusps}. For deep quenches like those considered in the previous subsections, we have shown that the Zak phase comes within some distance of the equilibrium Zak phase, $\mathcal{Z}^{PS}(t^*)\approx \mathcal{Z}^{eq}$. This is the first \textit{cusp}, or sharp peak or trough, in its non-equilibrium dynamics. The value of the Zak phase at the first cusp reflects the equilibrium topology of $\hat{H}_1'$. For deep quenches, all of the cusps (up to some late time) peak near the same value. We will say the cusps are of the same \textit{variety}. For brevity, we will call a cusp trivial if $\mathcal{Z}_{cusp} \approx 0$ (or $P_{cusp} \approx 0$), and topological if $|\mathcal{Z}_{cusp}| \approx 1$ (or $|P_{cusp}| \approx q a_0/2$). 

For shallow quenches close to the phase boundary, not all cusps are of the same variety. As we approach the phase boundary, cusps alternate between trivial and topological, and very close to the transition point, cusps at short times disappear. What \textit{survives} is that in all quenches (accessible with our numerical precision and system sizes considered) the topological character of the \textit{first} cusp aligns with the equilibrium topology of $\hat{H}_1'$. We can see this illustrated for the polarization in Fig.~\ref{fig:cusps}: as the topological phase boundary $J = d$ is approached from below, the initial topological cusp deforms and is replaced by a trivial cusp.

Given this knowledge, it is reasonable to try developing a criterion based on the Zak phase or the Resta polarization. For the Zak phase, the topology is determined by the first cusp which comes within some predefined tolerance of either $\mathcal{Z}^{PS} =  0$ or $|\mathcal{Z}^{PS}| =  1$. Practically, this criterion must exclude the temporal region around $t = 0$, as the Zak phase of the initial state is clearly trivial. Formally, let $\eta$ be our threshold for cusps, and let $c^{\mathcal{Z}}_{\eta}(\ket{\psi(t = 0)}, \hat{H}_1)$ be the function that finds the first cusp (within tolerance $\eta$) in the Zak phase and returns the character of the cusp,

\begin{align}
\label{eq:crit2}
c^{\mathcal{Z}}_{\eta} =
\begin{cases}
1 & \longleftrightarrow \hat{H}_1' \in \mathrm{TI}\\
0 & \longleftrightarrow \hat{H}_1' \in \mathrm{BI}
\end{cases},
\end{align}

where again the dependence on initial state and $\hat{H}_1$ is implicit. Astonishingly, for large enough systems (in our simulations $L = 24$ suffices) we are  able to completely recover the equilibrium phase diagram, even for parameter combinations near the topological phase boundary.

One can alternatively devise a similar criterion based on the state's Resta polarization, $P^{PS}$, which starts near the fractionalized value $|P^{PS}(t = 0)| \approx qa_0/4$. In analogy with the Zak phase, we let $c^{P}_{\eta}(\ket{\psi(t = 0)}, \hat{H}_1)$ be the function that finds the first cusp (within tolerance $\eta$ of $P^{PS} = 0$ or $|P^{PS}| = qa_0/2$) in the Resta polarization and returns the character of the cusp,

\begin{align}
\label{eq:crit3}
c^{P}_{\eta} =
\begin{cases}
1 & \longleftrightarrow \hat{H}_1' \in \mathrm{TI}\\
0 & \longleftrightarrow \hat{H}_1' \in \mathrm{BI}
\end{cases}.
\end{align}

Both criteria~(\ref{eq:crit2}) and~(\ref{eq:crit3}) are robust to the addition of small sublattice imbalance $\Delta_1 \neq 0$, and small interaction $V_1 \neq 0$. Additionally, we find numerically that they allow - without modification - dynamical recovery of the phase diagram, even when $\Delta_0$ is finite, so long as $\Delta_0 \gg \max{(J_0, d_0, \tau_0)}$. Fig.~\ref{fig:pol_cusp_phasediagram} exemplifies the generality of this dynamical recovery protocol based on the criterion~(\ref{eq:crit3}) for systems of $L = 10$ unit cells with $\Delta_1 = 0.1$ and $V_1 = 0.2$, where we have set $\tau_1 = 1$ as a consistent energy scale. Close to the phase boundary, we know that cusps are pushed to late times. To disregard finite size effects however, we set a maximum propagation time of $t_{max} = 20\tau_1$. Close to the phase boundary then, we expect cusps to occur at times $t > t_{max}$, leading to the region of uncertainty surrounding the equilibrium boundary line $J = d$ in Fig.~\ref{fig:pol_cusp_phasediagram}. For larger systems, this region is expected to shrink, and should vanish in the thermodynamic limit. 

We note that to our knowledge, nothing about the shape of these cusps is universal. Their presence, however, appears to be robust, and to allow for dynamical recovery of the equilibrium phase diagram of $\hat{H}_1'$ in a wide variety of circumstances. We expect similar cusp-like structure to emerge in other SPT systems under similar scenarios.

\begin{figure*}[htb!]
\includegraphics[width=0.98\textwidth]{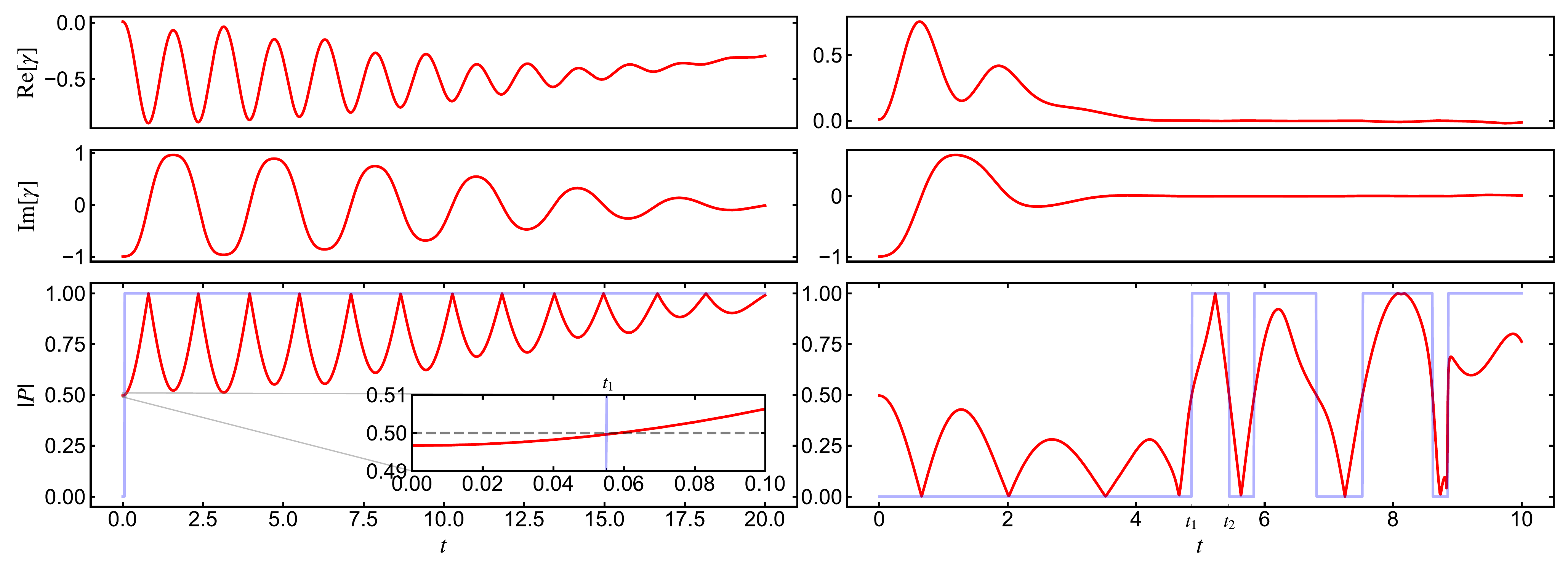} 
\caption{\label{fig:zte_quenches}Quenches from $\hat{H}_0\in$ OOBI to (left panel) $\hat{H}_1\in$ TI and (right panel) $\hat{H}_1\in$ BI. In both panels, red denotes symmetry broken state (i.e. $P^{PS}$) and blue denotes ZTE. In both cases, jumps in $P^{ZTE}$ are directly associated with zero crossings of $\mathrm{Re}[\gamma^{PS}]$. $P$ is in units of $q a_0/2$. Data shown for $L = 12$ unit cells with $V_1 = \Delta_1 = 0$ and $\tau_1 = 1$. For the left panel, $J_1 = 0.06$ and $d_1 = 1.0$, while for the right panel $J_1 = 1.3$ and $d_1 = 0.4$. The times of the first and second jumps in $P^{ZTE}$, as used in the stability analysis below, are labeled $t_1$ and $t_2$ in the lower panels. In the left panel, $t_2$, which is very large, is not shown.}
\end{figure*}

\section{Ensemble Signatures}
\label{sec:zte_sigs}
In the absence of symmetry breaking, the many body topology is necessarily pinned under unitary time evolution, whereas the single-particle topology can (under the right circumstances) undergo dynamical transitions. The main result of this section, as documented in rows 4 and 5 of Table~\ref{tab:quench_signatures}, is that spontaneous symmetry breaking flips the script. 

For quenches starting from SSB initial states, topological features are visible in the dynamics of the many-body topology (of the ZTE). If the post-quench Hamiltonian does not contain thermalizing interaction, then the single-particle topology (of the ZTE) is pinned while the many-body topology is not. These topological signatures manifest in the presence of `dynamical many-body topological' (DMBT) phases, as we will elucidate below. These DMBTs are more stable for quenches deeper into BI or TI. We clarify the sense in which these DMBTs are stable below. We also emphasize that, while we do report on (potentially universal) topological signatures in the post-quench dynamics, in contrast to our treatment of the pure state signatures above, we do not propose any explicit protocol for recovering the equilibrium phase diagram. If such an explicit protocol exists, we leave its investigation as a future research direction.

To understand ensemble topological signatures out of equilibrium, we employ many-body machinery. We assume that $\hat{H}_0$ spontaneously breaks chiral symmetry, resulting in two degenerate ground states. These states have large effective $\pm \Delta$, resulting in sublattice occupation predominantly on orbital \textit{a}~(\textit{b}), respectively. We emphasize, however, that these results only hold for \textit{effective} $\Delta_0$, as the properties we take advantage of disintegrate for $\Delta_0 \neq 0$ as well as for $\Delta_1 \neq 0$. Both $\hat{H}_0$ \textit{and} $\hat{H}_1$ must respect the SPT protecting symmetries. We further stress that while $\hat{H}_0$ and $\hat{H}_1$ obey the SPT symmetries, the initial states do not. However, as we will elucidate below, the ZTE does disclose, on differing timescales, the buried topology of $\hat{H}_0$, and the ground state topology of $\hat{H}_1'$. Moreover, these timescales generically differ from the timescales unearthed by our single-particle analysis in the previous section.

\subsection{Setup}
\label{subsec:zte_setup}
Here, we denote the two ground states by $\ket{\psi_0^{A(B)}}$. Intuitively, one can view $\ket{\psi_0^A}$ as a dressed version of the $V\rightarrow \infty$ eigenstate $\ket{A} = \prod_j  \hat{a}^{\dag}_j \ket{\emptyset}$ where $\ket{\emptyset}$ is the vacuum state, i.e. $\langle \psi_0^A|A \rangle \approx 1$, and similarly for $\ket{\psi_0^B}$. The ZTE at time $t = 0$ is comprised equally (as a result of energy degeneracy) of both ground states.

Upon quenching with $\hat{H}_1$, the two (initially order-obstructed) states evolve according to 
\begin{align}
    \ket{\psi^{A(B)}(t)} = \mathrm{exp}(-i \hat{H}_1 t)\ket{\psi^{A(B)}_0},
\end{align}

and the many-body density matrix retains the same weightings, due to $\hat{H}_1$ respecting sublattice symmetry. Thus, Eq.~(\ref{eq:rho_zte}) holds for all pre-quench and post-quench times.

\subsubsection{Ensemble Zak Phase}
\label{subsec:zte_zak}
While $\ket{\psi_0^{A(B)}}$ break sublattice symmetry, that $\hat{H}_0$ and $\hat{H}_1$ both respect sublattice symmetry allows us to gain insight into the Zak phase. Let $\mathcal{Z}^{A(B)}(t)$ be the instantaneous Zak phase at time $t$ for state $\ket{\psi^{A(B)}(t)}$. Because $\hat{H}_0$ respects sublattice symmetry, $\ket{\psi_0^{A(B)}}$ are particle-hole symmetric pair states. On the single-particle level, this is tantamount to interchanging lower and upper bands in the Bloch states. Because the Berry curvature integrated over all (both) bands must equal zero, $\mathcal{Z}^{A}(t = 0) = -\mathcal{Z}^B(t=0)$. Moreover, because $\hat{H}_1$ respects the protecting symmetries, this identity holds for all times $t$, or

\begin{align}
    \label{eq:zak_symm}
    & & \mathcal{Z}^{A}(t) = -\mathcal{Z}^B(t) \, & &\forall t,
\end{align}
implying that the same topological information is contained by both symmetry broken states. 

However, because the ZTE respects the protecting symmetries, so too does the SPDM, and consequently the Zak phase of the ensemble. This means that if $V_1 = 0$, the Zak phase of the ensemble is pinned to its initial value, i.e.

\begin{align}
    \label{eq:zak_zte}
    V_1 = 0 & &\Longrightarrow & &\mathcal{Z}^{ZTE}(t) = \mathcal{Z}^{ZTE}(t = 0),
\end{align}

and cannot uncover the ground state topology of $\hat{H}_1'$.

\subsubsection{Ensemble Polarization}
\label{subsec:zte_pol}
The Resta polarization of the ensemble, $P^{ZTE}(t)$, on the other hand, is mobile, and \textit{can} uncover the topological character of $\hat{H}_1'$.

In order to understand the novel possibilities for the polarization made possible by SSB, we introduce the variable 

\begin{align}
\gamma = \mathrm{Tr}[\hat{\rho}\, \mathrm{exp}(\frac{i 2 \pi}{L a_0} \hat{X})],
\end{align}

which allows us to rewrite the Resta polarization as $P = \frac{q a_0}{2 \pi} \mathrm{Im} \ln \gamma \quad (\mathrm{mod} \,q a_0)$. For the individual symmetry broken states, we have $\gamma^A(t):= \bra{\psi^A(t)}\mathrm{exp}(\frac{i 2 \pi}{L a_0} \hat{X})\ket{\psi^A(t)}$, and similarly for $\gamma^B(t)$.

\begin{figure*}[htb!]
\includegraphics[width=0.95\textwidth]{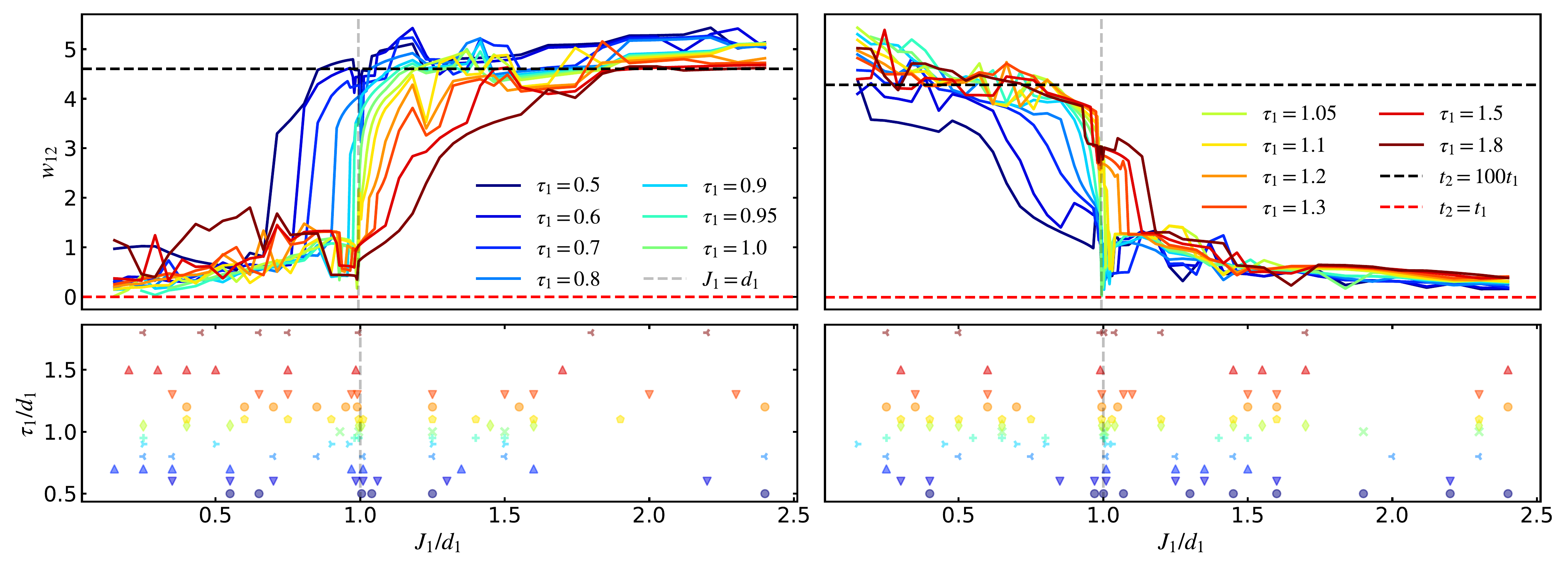} 
\caption{\label{fig:w_stability}Stability analysis for quenches from (left panel) $\hat{H}_0\in$ OOTI and (right panel) $\hat{H}_0\in$ OOBI, plotted for varied quench intercell hopping anisotropy $\tau_1$. Here, $t_1$ and $t_2$ are the times of the first and second jumps in $P^{ZTE}(t)$, and the dimensionless quantity $w_{12} = \log{(t_2/t_1)}$ tells us about the stability of the many-body dynamical phase. Upper panel illustrates the dependence on $J_1$ and $\tau_1$, with $t_2 \approx t_1$ implying stability for quenches within the same phase, and stability manifesting in $t_2 \gg t_1$ for quenches across equilibrium phase boundaries. Lower panel visualizes the positions of local minima in $w_{12}$ as a function of $J_1$ for various  $\tau_1$, as extracted from the data in the upper panel. The presence of minima at $J_1 \approx d_1$ for a wide range of $\tau_1$ suggests universality in the quench dynamics. Data shown for $L = 10$ unit cells with $V_1 = 0$ and $d_1 = 1$ to set the relative scale.}
\end{figure*}

In general $\gamma$ is complex valued, but in the non-interacting limit, $\mathrm{Im}[\gamma] = 0$ and the topology of a state is directly related to the sign of the real part according to
\begin{align}
P\big(\mathrm{Im} \, (\gamma) = 0\big) \quad = \quad 
\begin{cases}
      0 & \mathrm{sgn}\, \mathrm{Re}\,(\gamma) = 1 \\
      qa_0/2 & \mathrm{sgn}\, \mathrm{Re}\,(\gamma) = -1.
\end{cases}
\end{align}
The topology becomes ill-defined when both the real and imaginary parts of $\gamma$ are zero. In the opposite limit, $V \rightarrow \infty$, $\mathrm{Re}[\gamma] = 0$ and the polarization takes on the fractionalized value $P = \pm qa_0/4$. 

Whereas for the Zak phase, the symmetries of $\hat{H}_0$ and $\hat{H}_1$ manifested in Eq.~(\ref{eq:zak_symm}), for the polarization it results in 

\begin{align}
    \label{eq:pol_symm}
    \gamma^B(t) = (\gamma^A(t))^*,
\end{align}

which, when combined with~(\ref{eq:rho_zte}), yields

\begin{align}
    \label{eq:phi_zte}
    \gamma^{ZTE}(t) & = \big(\gamma^{A}(t) + \gamma^{B}(t)\big)/2 = \mathrm{Re}[\gamma^{A}(t)].
\end{align}

The Resta polarization of the ensemble then can be expressed as 
\begin{align}
    \label{eq:pol_zte}
    P^{ZTE}(t) \quad & =  \begin{cases}
      0 & |P^{PS}(t)| < q a_0/4  \\
      qa_0/2 & \mathrm{else}
\end{cases},
\end{align}

which we can see is intimately related to the polarization of the symmetry broken states. Importantly, $P^{\mathrm{ZTE}}(t)$ inherits the mobility from $P^{PS}(t)$, but unlike the symmetry broken polarization remains quantized. The initial value, $P^{ZTE}(t = 0)$, corresponds to the topology of the buried phase of the ground states of $\hat{H}_0$. However, upon quenching, it is possible for the order-suppressed polarization to undergo discrete jumps. Indeed, such jumps occur precisely when the many-body phase crosses $\mathrm{Re}[\gamma^{PS}] = 0$, which manifests in the symmetry broken states as $|P^{PS}(t)| = q a_0/4$. This relation between pure state and ensemble polarization is visualized in Fig.~\ref{fig:zte_quenches}, which shows exemplary quenches from the OOBI phase to both the TI (left panel) and BI (right panel). 

\subsection{Short Time Analysis}
\label{subsec:zte_shorttime}
In the previous subsection~\ref{subsec:zte_setup}, we showed in~(\ref{eq:pol_zte}) that $P^{ZTE}(t)$ is quantized and mobile. Moreover, we know that its initial value, $P^{ZTE}(t = 0)$, captures the order-obstructed topology of $\hat{H}_0$. In this section, we argue for the topological nature of the short-time behavior of this quantity. To do so, it is insightful to consider the dynamics of symmetry broken states, and then to relate this to the ZTE according to Eq.~(\ref{eq:pol_zte}). 

To develop intuition for the topological nature of these short-time topological signatures, it is informative to look at the short-time dynamics of the trivial product state, $\ket{A}=\prod_j  \hat{a}^{\dag}_j \ket{\emptyset}$. For simplicity, we first analyze the case $d_1 = \tau_1$, wherein there is no third neighbor hopping, and $V_1 = 0$ (i.e. no thermalizing interaction). The case $V_1 \neq 0$ is considered in the Appendix. For strong ordering in $\hat{H}_0$, ($V_0$ large), the dynamics of $\ket{A(t)}$ provides insight into the dynamics of $\ket{\psi^A(t)}$. At short times, we can Taylor expand the exponential $e^{-i \hat{H}_1 t} \approx \mathbb{1} - i\hat{H}_1 t + \mathcal{O}(\hat{H}^2_1 t^2)$. 

Let $|m, A\rangle = \hat{a}^{\dag}_m \ket{\emptyset}$ denote the single-particle state with occupation on the \textit{a} orbital of site $m$, and similarly for $|m, B\rangle$. In this notation, the product state is written as $\ket{A} = \prod_{j=1}^L |j, A\rangle$, and the action of the post-quench Hamiltonian $\hat{H}_1$ on one of the single-particle states is 

\begin{align}
    \hat{H}_1 |m, A\rangle = -J_1 |m, B\rangle - d_1 | m - 1, B\rangle,
    \label{eq:sp_Haction}
\end{align}

where site index is defined modulo $L$ for periodic systems. Moreover, because the initial state is a product state, we can decompose the action on the many-body state into a sum of contributions:

\begin{align}
    \hat{H}_1 \ket{A} = -\sum_{m = 1}^L \prod_{j \neq m} |j, A\rangle & \Big( J_1 |m, B\rangle  + d_1 | m - 1, B\rangle\Big).\nonumber
\end{align}

In the first order Taylor expansion, these terms can be interpreted as shifting occupation from \textit{a} to \textit{b} sublattice sites, where $|m,  B\rangle$ shifts to the right by half a unit cell, and $|m-1,  B\rangle$ shifts charge to the left. In an OBC system, charge would accumulate and build up on the one of the edges, with the relation $|J_1| \lessgtr |d_1| $ determining the direction of flow. In this case, charge is shifted to the right when $|J_1| > |d_1|$. If we instead consider starting from the product state $\ket{B}$, charge is shifted to \textit{a} orbitals in an analogous manner, and the direction of the shift is reversed. In PBC systems, there are no edges at which for charge to accumulate, but the direction of the shift tells us the direction of change in the Resta polarization.

While the initial (symmetry broken) states considered in our quenches are not identically product states, $\ket{\psi^A} \approx \ket{A}$, so the initial direction of flow for the Resta polarization is the same as for the product states. This directed movement in the short-time dynamics of the symmetry broken Resta polarization manifests in the ZTE in the presence of a jump in the quantized polarization when the buried topology of $\hat{H}_0$ differs from the ground state topology of $\hat{H}_1$. Additionally, the magnitude of $(|J_1| - |d_1|)$ tells us how quickly charge is shifted. This means that, the deeper we quench into the BI or TI phase, the faster we observe a jump in $P^{ZTE}(t)$.

\subsection{Stability Analysis}
\label{subsec:zte_stability}
In the short-time analysis above, we saw that in the absence of third-neighbor hopping ($d_1 = \tau_1$) and interaction ($V_1 = 0$), we can understand the short-time quench dynamics in terms of the competition between leftward and rightward shifting charge. Moreover, small $V_1 \neq 0$ does not qualitatively change this picture. However, as with pure state signatures, this simple picture breaks down upon adding third-neighbor hopping. More general understanding of the topological signatures, requires looking at the case $\tau_1 \neq d_1$.

\setlength{\tabcolsep}{15pt}

\begin{table*}[]
\captionof{table}{Post-Quench Dynamical Topological Signatures in interacting SSH Model~(\ref{eq:model})} \label{tab:quench_signatures}
\begin{center}
\begin{tabular}{l|lll}

Conditions                                                                   & $P$                         & $\mathcal{Z}$               & $\nu$              \\ \cline{1-4} 
\begin{tabular}[c]{@{}l@{}}No Interaction \quad \\ No SSB\end{tabular}           & Discrete \& Pinned          & Discrete \& Pinned          & Discrete \& Pinned \\ \cline{1-4} 

\begin{tabular}[c]{@{}l@{}}Thermalizing Interaction \quad \\ No SSB\end{tabular}              & Discrete \& Mobile          & \textbf{Discrete \& Mobile} & \textbf{Discrete \& Mobile} \\ \cline{1-4} 

\begin{tabular}[c]{@{}l@{}}Strong SSB (PS)\end{tabular}    & \textbf{Continuous}                  & \textbf{Continuous}                  & Discrete \& Pinned \\ \cline{1-4} 

\begin{tabular}[c]{@{}l@{}}No Interaction \quad \\ Strong SSB (ZTE)\end{tabular}     & \textbf{Discrete \& Mobile} & Discrete \& Pinned          & Discrete \& Pinned \\ \cline{1-4} 

\begin{tabular}[c]{@{}l@{}} Thermalizing Interaction \quad \\ Strong SSB (ZTE)\end{tabular}       & \textbf{Discrete \& Mobile}          & Discrete \& Mobile          & Discrete \& Mobile \\ \cline{1-4} 
\end{tabular}
\end{center}
\caption*{Comparison of single-particle and many-body topological signatures out of equilibrium for the interacting SSH model described by~(\ref{eq:model}). Boldface indicates that the `topological' marker's out-of-equilibrium dynamical behavior retains some notion of equilibrium topology. Pure state denotes the quench dynamics starting from a symmetry broken ground state. Zero-temperature thermal ensemble (ZTE) represents the equal mixture of symmetry broken ground states.}
\end{table*}

In equilibrium, topology is ill-defined at the point $\tau = 0$, implying that the magnitude of $\tau$ (relative to other terms in $\hat{H}$) relates to the stability of the notion of topology. Heuristically then, it makes sense for $\tau_1$ to influence the stability of any dynamical topological phases. In particular, we expect that for $\tau_1$ large (i.e. $\tau_1 > d_1$), the dynamical topological phase will be bolstered. For $\tau_1 < d_1$ on the other hand, the stability of the topological phase will be degraded. Regardless of $\tau_1$ however, we expect that the DMBTs are more stable for quenches further from the equilibrium topological boundary.

In order to consistently define a notion of dynamical stability for post-quench Hamiltonians with arbitrary system parameters (and information propagation velocity), we identify two times of interest. Namely, we take the times of the first and second jumps ($t_1$ and $t_2$ resp.) in $P^{ZTE}(t)$. For the sake of comparing timescales, we define~\footnote{We note that the ratio $t_2/t_1$ is dimensionless in its own right, and thus invariant to multiplicative factors applied to the entire quench Hamiltonian. One could just as well have worked with the bare ratio. We find its logarithm to align better with intuitive notions of stability.}

\begin{align}
    \label{eq:stability}
    w_{12} = \log{(t_2/t_1)}.
\end{align}

To see why~(\ref{eq:stability}) indicates the stability of the DMBT phase, we consider two types of quenches. 

For quenches in which the buried topology of $\hat{H}_0$ and the ground state topology of $\hat{H}_1$ align, $P^{ZTE}$ accurately reflects this topology up until $t_1$, at which time it jumps to the equilibrium value corresponding to the other phase. If $t_2$ is not much larger than $t_1$, then the dynamical Resta polarization assumes the `correct' topological value for a long time (relative to $t_2$), and we regard the system as very stable. In this case, $t_2 \approx t_1$ results in $w_{12} \approx 0$. Hence, for quenches within an (order obstructed) phase, $w_{12} \rightarrow 0$ indicates maximal stability.

For quenches that cross equilibrium topological phase transitions, (e.g. $\hat{H}_0 \in $ OOTI, $\hat{H}_1\in$BI), the speed of the transition (and therefore $t_1$) should depend on the depth of the quench. Such dynamical phases can be regarded as stable when $t_1$ is short and $t_2$ is long (relative to $t_1$), resulting in large $w_{12}$.

Thorough numerical simulations strongly support the interpretation of this as dynamical topological behavior. We find that in great generality, deeper quenches result in more \textit{stable} many-body dynamical topological phases. This is illustrated in Fig.~\ref{fig:w_stability}, where $J_1/d_1 \ll 1$ and $J_1/d_1 \gg 1$ result in very different qualitative behavior, regardless of $\hat{H}_0$. Moreover, we observe an intriguing crossover between these two regimes, which is (possibly universally) characterized, independent of $\tau_1$, by local minima in $w_{12}$ around the equilibrium topological phase boundary, in this case $J_1 = d_1$. 

As in the pure state discussion above, this generalized machinery for characterizing stability is unnecessary for $\tau_1 = d_1$. In this special case, it is sufficient to look at $t_1$, which should increase as $J_1$ approaches $d_1$, tending toward arbitrarily late times in the limit $J_1 \rightarrow d_1$.

As we saw above, the dynamical topological behavior of the Resta polarization following quenches from SSB initial states is genuinely many-body, owing to constraints imposed upon the many-body density matrix. Remarkably, this behavior does not require thermalizing interaction.

To recapitulate, starting from an ensemble of spontaneous symmetry breaking states of an SPT pre-quench Hamiltonian, and evolving with an SPT post-quench Hamiltonian, we have shown that a representative many-body topological invariant, namely the Resta polarization, becomes mobile while retaining quantization. At the beginning of the quench, this quantity is guaranteed to reveal the buried topology of the pre-quench Hamiltonian. For short times, in the absence of third-neighbor hopping, we have argued that the dynamics of the polarization should be dominated by the equilibrium topology of the post-quench Hamiltonian. More generally, we have defined a metric for stability of the dynamical signatures, and have set forth strong numerical evidence that this notion of stability strongly correlates with distance (of the post-quench Hamiltonian) from the equilibrium topological phase boundary. Finally, and more speculatively, we have identified potentially universal features in the stability metric which coincide with the equilibrium topological phase boundary. Taken together, these findings lead us to define dynamical many-body topology (DMBT), crucially enabled by spontaneous symmetry breaking.

\section{Discussion}

In this paper we have demonstrated that, counterintuitively, signatures of topology in SPT systems can survive in the non-equilibrium dynamics even when a protecting symmetry is broken by an obstructing order. Studying an interacting variant of the SSH model, we have identified such topological signatures in the dynamics of the pure state, and of the zero-temperature ensemble, in which the obstructing order is explicitly suppressed.  These signatures are summarized in Table~\ref{tab:quench_signatures}.

For pure states, we find cusps in the post-quench dynamics of representative single-particle and many-body topological signatures. The value of the first such cusp reflects the ground state topology of the post-quench Hamiltonian, allowing us to dynamically recover the equilibrium phase diagram of the post-quench Hamiltonian. Moreover, when small interacting and symmetry breaking terms are added to the post-quench Hamiltonian, we are still able to dynamically recover the equilibrium phase diagram of the symmetry respecting, single-particle portion of the post-quench Hamiltonian.

For zero-temperature ensembles, we have shown that spontaneous symmetry breaking in conjunction with non-equilibrium dynamics can give rise to what we dub dynamical many-body topology (DMBT). In all previously investigated scenarios, to our knowledge, the many-body topology is pinned under unitary time evolution. Therefore, previous discussion of dynamical topological behavior was inherently of single-particle nature. We have presented an alternative scenario in which the single-particle topology is pinned while the many-body topology is mobile. 

While the numerics and some of the functional forms for the Zak phase and Resta polarization were specific to the model investigated, we stress that the majority of our results should apply, with appropriate modification, to order obstructed SPT systems out of equilibrium more broadly. We do not present a proof for the general validity of our topological signatures. However, by considering extended-range interactions and third-neighbor hopping, we believe we have separated the system-specific results from the generic conclusions. We also note that none of the theoretical arguments utilized the dimension of our system. The basic phenomena documented above should extend to SPT systems in higher dimensions.


\begin{acknowledgments}
{\textit{Acknowledgements. --} }We thank Jan C. Budich for his guidance and helpful discussion. We also thank the Stanford Research Computing Center for providing computational resources. Supported by the U.S. Department of Energy (DOE), Office of Basic Energy Sciences, Division of Materials Sciences and Engineering, under contract DE-AC02-76SF00515. M. S. thanks the Alexander von Humboldt Foundation for its support with a Feodor Lynen scholarship. 
\end{acknowledgments}



\section*{Appendix A: Pure State Signatures}
This appendix contains details regarding Sec.~\ref{sec:pure_state_sigs} in the main text, as well as supporting results. First, we explicitly compute the bound~(\ref{eq:zak_dyn_II_bound}) in the main text. Then we numerically demonstrate the robustness of the pure state protocols~((\ref{eq:crit2}) and~(\ref{eq:crit3}) in the main text) against sublattice imbalance which breaks the protecting chiral symmetry. Finally, we extend the discussion to initial states with finite $\Delta_0$ and demonstrate that for $\Delta_0$ sufficiently large (but still finite), the arguments in the main text apply with only slight modification.

\begin{figure*}[htb!]
\includegraphics[width=0.97\textwidth]{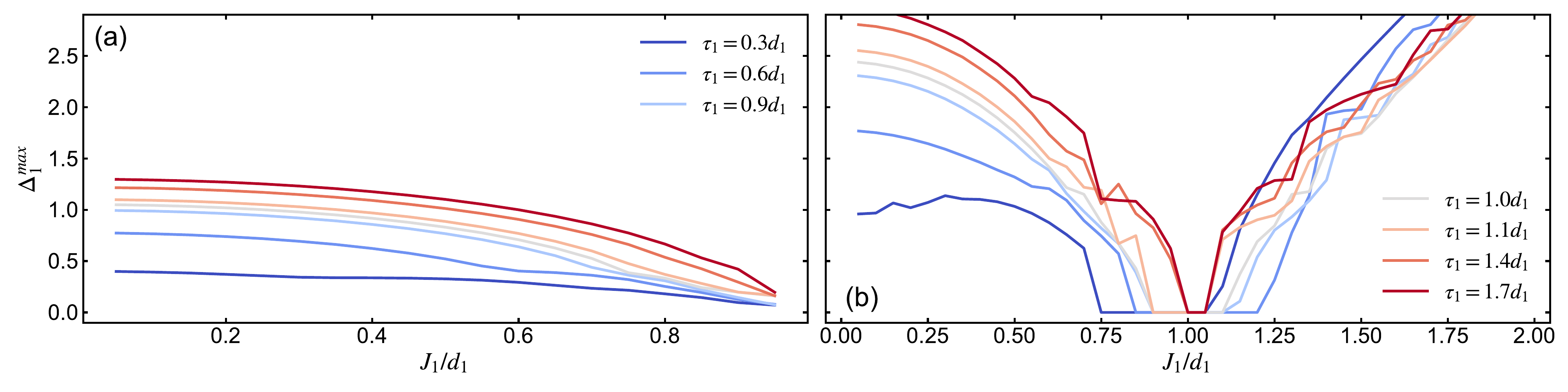} 
\caption{\label{fig:cusp_dlt1_rob}Demonstration of robustness of successful pure state protocols for dynamical recovery of equilibrium topology from quench dynamics of (a) Zak phase~(\ref{eq:crit2}), and (b) Resta polarization~(\ref{eq:crit3}). In both cases, the system starts as a $\Delta \rightarrow \infty$ eigenstate, and evolves according to quench Hamiltonian $\hat{H}_1$, with $V_1 = 0$. $\Delta_1^{max}$ represents the largest $\Delta_1$ such that the protocol is successful. (a) Data shown for $L = 32$, obtained from tight binding calculations. For $\Delta_1 > \Delta_1^{max}$, the cusps are trivial. Values only shown for $J_1 > d_1$, because when $J_1 < d_1$, the system is trivial. Addition of any sublattice imbalance, which explicitly breaks the protecting chiral symmetry, results in another trivial Hamiltonian with cusps at $\mathcal{Z}^{PS} = 0$, so there is no $\Delta_1^{max}$. A tolerance of $\eta_{\mathcal{Z}}$ was used. (b) Resta polarization for system of size $L = 8$ computed in ED. Unlike the Zak phase, when $\Delta_1 > \Delta_1^{max}$, $P^{PS}$ exhibits no cusps at all. As a result, it makes sense to define $\Delta_1^{max}$ for $J_1 > d_1$ as well as $J_1 < d_1$. A tolerance of $\eta_{\mathcal{P}}$ was used. In both (a) and (b), robustness increases with distance from the phase boundary (equivalently depth of quench). This general trend holds regardless of $\tau_1$. Additionally, $\tau_1 > d_1$ bolsters the TI, whereas $\tau_1 < d_1$ bolsters the BI phase. This mirrors the results shown in Fig.~\ref{fig:w_stability} in the main text. All quenches are stopped at $t_{max} = 20$.}
\end{figure*}

\subsection*{Derivation of Eq.~(\ref{eq:zak_dyn_II_bound})}
Let \begin{align}
    I_1 = \oint{dk \cos{(k)} \nabla_k \phi_k^{SSH}}.
\end{align}

This has the form of $\int{udv}$ with $u = \cos{(k)}$ and $dv = \nabla_k \phi_k^{SSH}dk$. By integration by parts, this equals $\int{vdu}$, or $I_1 = \oint{dk(-\sin{(k)})\phi^{SSH}_k}$, where the boundary contribution $uv$ vanishes because the path is a closed loop.

$J \gg d$, so $h^x_k > 0$ for all $k$, and $\arg$ simplifies to $\phi_k^{SSH} = \arctan{(h^y_k/h^x_k)}$, and we have $I_1 = \oint{dk (-\sin{(k)}) \arctan{(h^y_k/h^x_k)}}$. $|\sin{(z)}| \leq 1 \, \forall z$, and $|\arctan{(z)}| \leq \pi \, \forall z$, so 

\begin{align*}
    |I_1| = & |\oint{dk \cos{(k)} \nabla_k \phi_k^{SSH}}| \\
    = & |\oint{dk \sin{(k)}\phi^{SSH}_k}| \\
    \leq & \oint{dk |\sin{(k)}| |\arctan{(h^y_k/h^x_k)}|} \\
    \leq & \int_{\pi}^{\pi}{\pi dk} \\
    = & 2 \pi^2.
\end{align*}

Multiplying this result by $|-\delta/4|$, we arrive at Eq.~(\ref{eq:zak_dyn_II_bound}).

\subsection*{Robustness to $\Delta_1$}
In Sec.~\ref{subsec:ps_dlt1} in the main text, we showed that for $\Delta_1 \ll d_1$, the Zak phase at time $t^*$ is independent of $\Delta_1$, while $t^*$ itself only weakly depends on the small quantity $\chi = \Delta_1/d_1 \ll 1$. The criteria~((\ref{eq:crit2}) and~(\ref{eq:crit3}) we eventually found to allow for dynamical recovery of equilibrium topology were \textit{guided} by our intuitions of robustness, but were not directly related to $t^*$ in the general case. Moreover, the dynamical recovery protocol employed in the main text (Fig.~\ref{fig:pol_cusp_phasediagram}) was performed using $\Delta_1 = 0$. Here, we explicitly demonstrate (and test the limits of) robustness with respect to $\Delta_1$ by identifying the largest $\Delta_1$ (which we call $\Delta_1^{max}$) for a given $J_1$, $d_1$, $\tau_1$ (with $V_1 = 0$) for which the dynamical recovery protocol successfully determines the equilibrium topology of $\hat{H}_1'$. The results, shown in Fig.~\ref{fig:cusp_dlt1_rob}, demonstrate that for deep quenches, the topological signatures remarkably persist for large $\Delta_1$ (even for $\Delta_1 >d_1$). As expected, $\Delta_1^{max}$ increases the farther the quench is from the equilibrium phase boundary. Finally, as in the main text we see that $\tau_1$ large stabilizes the TI phase, making for greater tolerance of $\Delta_1$.

\subsection*{Single Particle Treatment for finite $\Delta_0$}
In the main text, we only considered cases in which $\hat{H}_0 = \hat{\Delta}$. From the standpoint of a more generic Hamiltonian $\hat{H}_0 = \hat{H}^{SSH}(J, d, \tau, v) + \hat{\Delta}$, this represents the limit $\Delta_0 \rightarrow \infty$. More broadly, we are interested in the case of finite $\Delta_0$.

Here we show that the physics characterized in the main text remains essentially unchanged for large initial symmetry breaking ($\Delta_0$ large). We will use the subscript `$0$' or `$1$' to distinguish parameters in the ground state and quench Hamiltonian respectively. We will focus on the more analytically tractable case $h_0^x = h_1^x$, and $h_0^y = h_1^y$, which still illustrates the effect of finite $\Delta_0$. This allows us to drop the subscript for $J$, $d$, and $\tau$. As in case III in the main text, we will take $\delta = d/J = \tau/d \ll 1$, $\chi = \Delta_1/J \ll 1$, but now we introduce the additional small parameter $\alpha = J/\Delta_0 \ll 1$, which enforces strong symmetry breaking ($\lim_{\Delta_0\rightarrow \infty}\alpha = 0$). We will expand up to second order in the small quantities $\alpha$, $\delta$, and $\chi$. We will also find it insightful to introduce the quantity $r_{\Delta} = \alpha \chi = \frac{\Delta_1}{\Delta_0}$, which encapsulates the relative strength of symmetry breaking in $\hat{H}_0$ and $\hat{H}_1$.

We proceed by generalizing Eq.~(\ref{eq:u_dlt})~-~(\ref{eq:rho_k_t_constraint}) in the main text. We parameterize the Hamiltonian vector on the Bloch sphere as 

\begin{align}
\label{eq:finitedlt0_bloch}
    \vec{\hat{h}}(k) = (\sin{\Theta_k}\cos{\Phi_k}, \sin{\Theta_k}\sin{\Phi_k}, \cos{\Theta_k}),
\end{align}

where 
\begin{align}
    \label{eq:finitedlt0_ThetaPhi}
    \Theta_k &= \arccos{(h^z_k/h_k)}, \\
    \Phi_k &= \arctan{(h^y_k/h^x_k)},
\end{align}

represent the azimuthal and polar angles respectively. The conditions $h_0^x = h_1^x$, and $h_0^y = h_1^y$ here simplify matters by rendering the polar angle constant ($\Phi_k = \Phi^0_k = \Phi_1$). Dropping the explicit momentum dependence, the initial state (eigenstate of $H_0$) takes the form

\begin{align}
    \label{eq:finitedlt0_u0}
    \ket{\mathscr{u}_k(t = 0)} =  \begin{pmatrix} \sin{(\frac{\Theta_0}{2})} \\ -\cos{(\frac{\Theta_0}{2})e^{i\Phi}} \end{pmatrix},
\end{align}

and evolves in time according to

\begin{align}
     \label{eq:finitedlt0_ut}
    \ket{\mathscr{u}_k(t)} =&  \cos{(h_1t)} \ket{\mathscr{u}_k(t = 0)} \nonumber \\
    &+i\sin{(h_1t)}\begin{pmatrix} \sin{(\Theta_1 - \frac{\Theta_0}{2})} \\ -\cos{(\Theta_1 - \frac{\Theta_0}{2})e^{i\Phi}} \end{pmatrix}.
\end{align}

In terms of $\alpha$, $h_0 \approx \Delta_0 (1 + \alpha^2/2)$ and $\hat{h}_0^z \approx 1 - \alpha^2$. The angle $\Delta\Theta = \Theta_1 - \Theta_0/2$ contains the azimuthal dependence of the initial state and the quench Hamiltonian. $\Theta_0\approx \alpha$ reflects the finitude of $\Delta_0$, with larger $\Theta_0$ for smaller $\Delta_0$. $\Theta_1 \approx \frac{\pi}{2} - \chi(1 - \delta \cos{(k)})$ reflects the symmetry breaking in $H_1$. The difference takes the form 

\begin{align}
\label{eq:DeltaTheta}
    \Delta\Theta \approx \frac{\pi}{2} - \chi(1 - \delta \cos{(k)}) - \frac{\alpha}{2}.
\end{align}

In the well-behaved limit $\Delta_0 \rightarrow \infty$, $\Delta_1 \rightarrow 0$ (ground state is symmetry breaking eigenstate, quench Hamiltonian is symmetry respecting), we have $\Delta \Theta = \frac{\pi}{2}$. From (\ref{eq:DeltaTheta}, we can interpret $\alpha$ and $\chi$ as small contributions to a (parameterically small) shift away from the ideal case. Solving Eq.~(\ref{eq:rho_k_t_constraint}), we arrive at 

\begin{align*}
    t^* = \frac{\pi}{4J}\big[ 1 - \frac{3}{4}\delta^2 + (\frac{4}{\pi} - 1)\frac{\chi^2}{2} + \frac{2}{\pi}r_{\Delta}\big],
\end{align*}

which includes a new contribution coming from the non-vanishing ratio of symmetry breaking in $\hat{H}_0$ and $\hat{H}_1$. As was the case with $\Delta_1 \neq 0$, we see that $\alpha \neq 0$ shifts the critical time $t^*$, but its role in the expectation values of observables cancels out. This means that the single-particle dynamical topological classification scheme introduced in the main text should (potentially with slight modification) also work starting from initial states which exhibit only moderate to strong (but not ideal) symmetry breaking. Numerically, we find that this intuition shines through in the Zak phase, which shows negligible (sub-polynomial) dependence on $\alpha$.

\subsection*{Incorrect Generalization to $\tau \neq d$}

In Sec.~\ref{subsec:ps_phasediagram} in the main text, we provide an explicit procotol - based on intuition from the simple case $\tau = d$ and $V = 0$ - for dynamical recovery of equilibrium topology. We acknowledge that the basic intuition could also lead one to consider alternative, potentially appealing generalizations. Here, we detail one particularly appealing generalization, and demonstrate the shortcomings of this method.

When we made the assumption of quenching deep into the BI or TI phase, we saw that at time $t^*$, the deviation in the Zak phase, $\mathcal{Z}^{dyn}(t^*)$, from the equilibrium topological invariant $\mathcal{Z}^{eq}$ was parameterically small (e.g.~(\ref{eq:zak_dyn_I})) according to the depth of the quench, as was the deviation in $\langle \hat{S}^z_k(t^*)\rangle$ from zero. This fortuitous fact had to do with $\ket{\mathscr{u}(t)}$ approximately being an eigenstate of $H_1'$ at time $t^*$. In the absence of a dynamical quantum phase transition (DQPT), the system is expected, at some time $\tilde{t}$ in the vicinity of $t^*$, (i.e. $\tilde{t} \approx t^*$) to cross $\langle \hat{S}^z_k(\,\tilde{t}\,)\rangle = 0$. 

For quenches deep into TI or BI, $\ket{\mathscr{u}(\,\tilde{t}\,)}$ is also approximately an eigenstate of $H_1'$, leading to similar behavior in the Zak phase as at $t^*$. Here we also introduce $\ket{\psi(t)}$ as the many-body state corresponding single-particle state $\ket{\mathscr{u}(t)}$, and note that $\ket{\mathscr{u}(\,\tilde{t}\,)}$ approximately being an eigenstate of $H_1'$ implies $\ket{\mathscr{u}(\,\tilde{t}\,)}$ is an approximate eigenstate of $\hat{H}_1'$. This time $\tilde{t}$ also allows for more equitable comparison of quenches near and far from the equilibrium boundary. In particular, we define $\tilde{t}(\ket{\psi(t = 0)}, \hat{H}_1)$ to be the first time at which the pseudospin $z$-component crosses zero for state $\ket{\psi}$ evolving according to Hamiltonian $\hat{H}_1$. 

We are implicitly assuming in this discussion (and this criterion) that $\ket{\psi(t = 0)}$ is sufficiently symmetry broken (i.e. $\Delta_0$ or $V_0$ is large compared to $J_0$, $d_0$, and $\tau_0$). Importantly, we see that this provides a consistent time-scale which is agnostic to system parameters. Moreover, this allows generalization to interacting systems. While we don't expect $\mathcal{Z}(\tilde{t})$ to line up exactly with $\mathcal{Z}^{eq}$, our hope, guided by the limiting cases detailed above, is that the value of the Zak phase at this time tells us about the topology of $\hat{H}_1'$. Recognizing that $0 \leq |\mathcal{Z}^{PS}| \leq 1$, as a first (naive) criterion, we designate

\begin{align}
\label{eq:crit1}
    \mathcal{Z}^{PS}(\,\tilde{t}\,) > \frac{1}{2} & \longleftrightarrow \hat{H}_1' \in \mathrm{TI} \\
    \mathcal{Z}^{PS}(\,\tilde{t}\,) < \frac{1}{2} & \longleftrightarrow \hat{H}_1' \in \mathrm{BI},
\end{align}

\begin{figure}[htb!]
\includegraphics[width=0.48\textwidth]{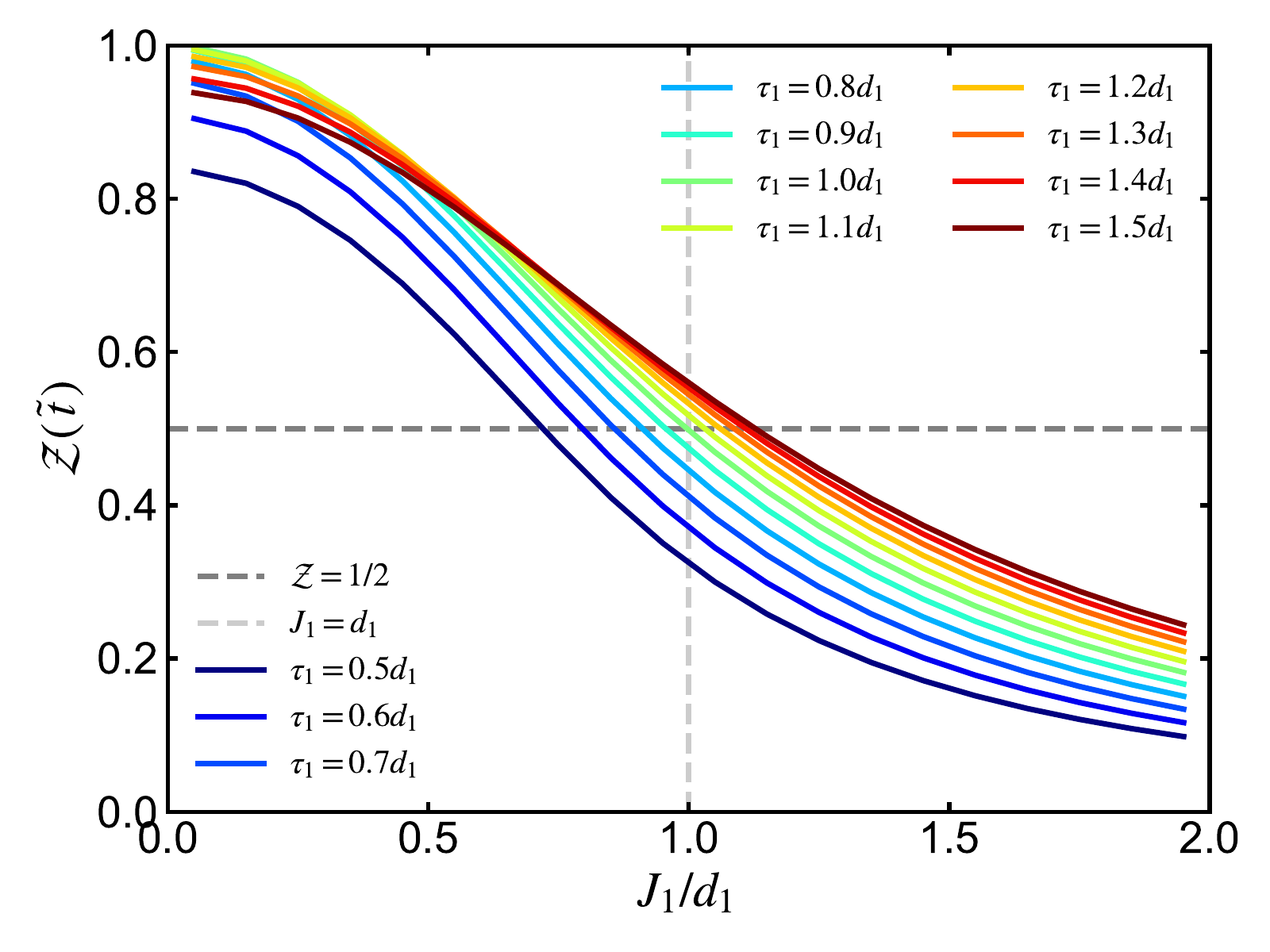} 
\caption{\label{fig:crit1_breakdown}The pure state Zak phase at time $\tilde{t}$ for quench Hamiltonians with variable third-neighbor hopping. The first non-equilibrium pure state criterion (Eq~(\ref{eq:crit1})) diagnoses all Hamiltonians with $\mathcal{Z}(\,\tilde{t}\,) > \frac{1}{2}$ (above the horizontal grey dashed line) as topological. The true equilibrium topological condition, $J_1 < d_1$, (to the left of the vertical dashed-dotted line), as topological. While these two conditions align for $\tau_1 = d_1$, there is a mismatch for $\tau_1 \neq d_1$. Data shown for $L= 24$ unit cells with $\Delta_1 = V_1 = 0$.}
\end{figure}

where the dependence of $\tilde{t}$ on $\hat{H}_1$ and on $\ket{\psi(t=0)}$ are left implicit. This criterion~(\ref{eq:crit1}) actually does hold true for arbitrary $J_1$ and $d_1$, so long as there is no third-neighbor hopping ($\tau_1 = d_1$). In other words, it applies to the canonical SSH model, with $\mathcal{Z}(\,\tilde{t}\,)$ changing monotonically with the distance to equilibrium topological boundary. For $\tau_1 \neq d_1$ however, the criterion is imperfect, as illustrated in  as Fig~\ref{fig:crit1_breakdown}. In particular, while the criterion makes correct predictions for large swaths of parameter space, it predicts the incorrect equilibrium topology in the upper left and lower right quadrants of Fig.~\ref{fig:crit1_breakdown}.
\section*{Appendix B: Ensemble Signatures}
\begin{figure*}[htb!]
\includegraphics[width=0.97\textwidth]{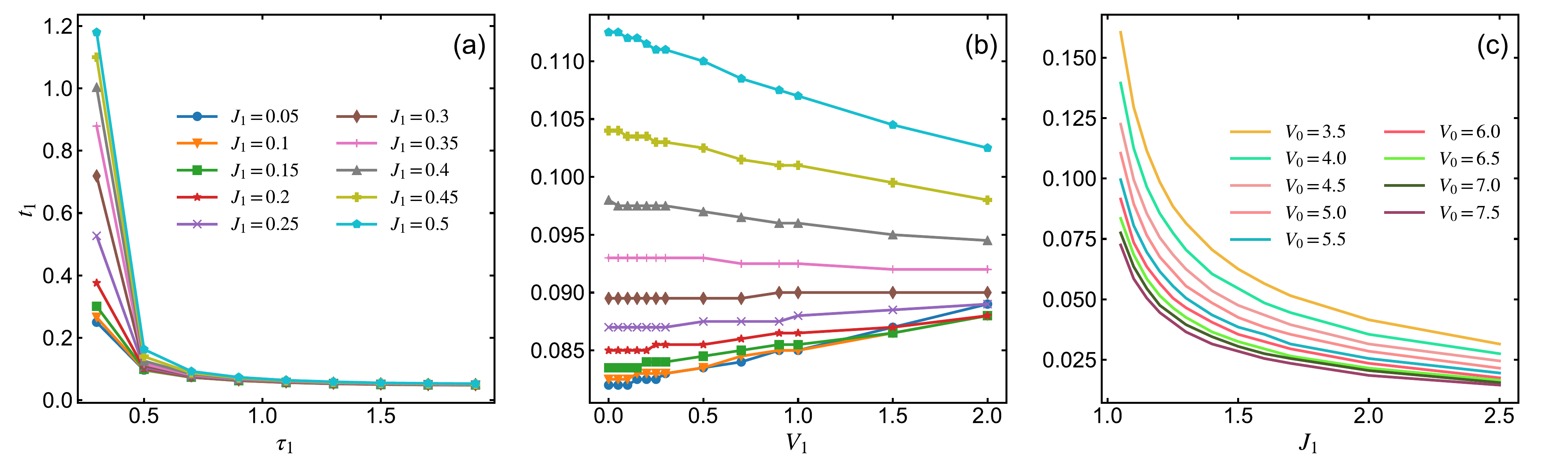} 
\caption{\label{fig:dmrg_param_dep}Dependence of $t_1$ on various system parameters for quenches which cross an equilibrium topological phase boundary. (a) and (b) show quenches from OOBI $\rightarrow$ TI, whereas (c) shows quenches from OOTI $\rightarrow$ BI. (a): Influence of third-neighbor hopping on first critical time. $t_1$ clearly decreases for deeper quenches (smaller $J_1$). This trend holds for all values of $\tau_1$. As $\tau_1 \rightarrow 0$, topology becomes ill-defined, which is reflected in the drastic increase in $t_1$, likely decreasing the stability of the dynamical topological phase. (b): Influence of quench interaction $V_1$ on $t_1$. While the exact time of the jump in $P^{ZTE}$ changes with $V_1$, the behavior is qualitatively the same as for $V_1 = 0$, demonstrating robustness of the dynamical topological signatures. (c) Influence of initial interaction $V_0$ on $t_1$. Larger $V_0$ (relative to other parameters in $\hat{H}_0$) brings the initial state closer to the trivial product state. The initial symmetry broken Resta polarization, $P^{PS}$, thus starts (at $t = 0$) closer to $|P^{PS}| = qa_0/4$, and has a shorter distance to travel in Hilbert space before undergoing a jump in $P^{ZTE}$.  Systems of size $L = 64$ unit cells are considered in all cases. For all quenches, $d_0 = 0.22$, $\tau_0 = 0.3$, and $d_1 = 1.0$. For (a) and (b), $J_0 = 0.75$ and $V_0 = 5.0$, while $V_1 = 0$ for (a) and $\tau_1 = 0.6$ for (b). In (c), we have set $J_0 = 0.1$, $V_0 = 5$, and $tau_1 = 0.9$ for each quench.}
\end{figure*}

In the main text, we presented numerical results for quenches obtained using Exact Diagonalization (ED) and tight binding (TB) methods. ED calculations were performed using the QuSpin library~\cite{quspin1, quspin2}. TB calculations were performed using the Python Tight Binding package~\cite{pythtb}.

The Resta polarization, which is a genuine many-body quantity, cannot be extracted from TB formalism, which severely limited accessible system sizes for our complete stability analysis. Restricting our attention to the half-filled case and utilizing translation invariance to work in the ground state momentum sector, we were only able to perform quenches with $L=12$ (24 orbitals). To rule out the possibility that our findings are due to finite-size effects, of course we would like to probe larger systems using alternate methods where possible.

In this appendix, we employ Density Matrix Renormalization Group (DMRG) techniques to investigate the role of various system parameters (beyond the simple case considered in Sec.~\ref{subsec:zte_shorttime} in the main text) in the dynamical topological behavior visible in the zero temperature ensemble. In particular, we discuss the dependence of $t_1$ (the time of the first jump in $P^{ZTE}$) on $V_0$, $V_1$, and $\tau_1$. The results are shown in Fig.~\ref{fig:dmrg_param_dep}. DMRG calculations are performed using the ITensor library \cite{itensor}.

DMRG techniques do not entirely eradicate the issues plaguing interacting quantum systems, but they do allow us to corroborate the findings in the main text. In particular, DMRG calculations allow us to consider much larger systems at the expense of poor scaling (due to the build-up of entanglement entropy) with the propagation time. More precisely, we use DMRG to generate approximate ground states of $\hat{H}_0$ in matrix product state (MPS) representation, and then propagate forward in time by approximately exponentiating the matrix product operator (MPO) representation of $\hat{H}_1$. Entanglement entropy across bipartitions of the system grows linearly with time, leading to exponential growth of the requisite bond dimension for the MPS. Moreover, the scaling is made quadratically worse by the fact that we must deal with periodic systems (which have twice the entanglement entropy as do open boundary condition (OBC) systems). Together, these considerations conflate to prohibit full-scale DMRG analysis of many-body topological quench dynamics. 

Nevertheless, for quenches which cross an equilibrium topological phase boundary (e.g. $\hat{H}_0 \in $ OOTI, $\hat{H}_1 \in$ BI), we expect the first jump (at time $t_1$) in the ensemble polarization $P^{ZTE}$ to occur at a relatively early time, especially for deep quenches. Accordingly, we employ DMRG for systems of $L = 64$ unit cells to investigate the dependence of $t_1$ on system parameters, and to bolster the story conveyed in the main text.

\paragraph{The role of $\tau_1$}---
As we saw in Fig.~\ref{fig:w_stability} in the main text, $\tau_1>d_1$ favors the TI phase, while $\tau_1 < d_1$ favors the BI phase. As $\tau_1 \rightarrow 0$, the dynamical topological phase is completely destabilized, owing to the topology being ill-defined at the point $\tau = 0$. Similar behavior is visible in $t_1$, which exhibits two qualitatively very different regimes for $\tau_1 \lessgtr d_1$, as seen in Fig.~\ref{fig:dmrg_param_dep}(a).

\paragraph{The role of $V_1$}---
For the symmetry breaking product state $\ket{A}$, $\hat{V}\ket{A} = 0$ because $\langle A | \hat{n}^b_j|A\rangle = 0 \, \, \forall j$ (and similarly for $\ket{B}$ with $\hat{n}^a_j$. Thus, for strongly symmetry broken initial states (such as the ones we consider in the main text), weak interaction $V_1$ should have very little effect on the dynamics. To first order, we can approximate $\hat{H}_1$ for small $V_1$ with an effective non-interacting Hartree-Fock Hamiltonian, where $d_1$ and $\tau_1$ are both renormalized to $\tilde{d}_1 > d_1$, and $\tilde{\tau}_1 > \tau_1$. As we saw in the previous paragraph, larger $\tau_1$ favors the topological phase. Additionally, larger $d_1$ for the same $J_1$ behaves effectively like a deeper quench (smaller $J_1$). This trend is apparent in the top few curves in Fig.~\ref{fig:dmrg_param_dep}(b), for $t_1$ decreases as $V_1$ increases. For the lower curves ($J_1\leq 0.25$) which show upward movement in $t_1$, this only happens for larger $V_1$, where other effects dominate. 

\paragraph{The role of $V_0$}---
The interaction term in $\hat{H}_0$ induces spontaneous symmetry breaking, and energetically favors CDW order. The eigenstates of $\hat{V}_0$ satisfy $\langle \hat{S}^z\rangle = 1$ and $|P^{PS}|- q a_0/4=0$ identically. Competition between interaction and hopping terms for finite $V$ results in $\langle \hat{S}^z\rangle < 1$ and $||P^{PS}| - q a_0/4|>0$. For larger $V$, both sublattice imbalance and polarization approach their $V\rightarrow \infty$ values. The closer $P^{PS}$ is to $P^{PS}=\pm q a_0/4$, the less shifting of charge needed to reach (and cross) $|P^{PS}|= qa_0/4$, and for a corresponding jump in $P^{ZTE}$ to occur. Thus, we should expect that $V_0$ strongly influences the timescales associated with the dynamical many-body topological transitions. This is borne out in numerical experiments, as depicted in Fig.~\ref{fig:dmrg_param_dep}(c).


\begin{thebibliography}{61}%
\makeatletter
\providecommand \@ifxundefined [1]{%
 \@ifx{#1\undefined}
}%
\providecommand \@ifnum [1]{%
 \ifnum #1\expandafter \@firstoftwo
 \else \expandafter \@secondoftwo
 \fi
}%
\providecommand \@ifx [1]{%
 \ifx #1\expandafter \@firstoftwo
 \else \expandafter \@secondoftwo
 \fi
}%
\providecommand \natexlab [1]{#1}%
\providecommand \enquote  [1]{``#1''}%
\providecommand \bibnamefont  [1]{#1}%
\providecommand \bibfnamefont [1]{#1}%
\providecommand \citenamefont [1]{#1}%
\providecommand \href@noop [0]{\@secondoftwo}%
\providecommand \href [0]{\begingroup \@sanitize@url \@href}%
\providecommand \@href[1]{\@@startlink{#1}\@@href}%
\providecommand \@@href[1]{\endgroup#1\@@endlink}%
\providecommand \@sanitize@url [0]{\catcode `\\12\catcode `\$12\catcode
  `\&12\catcode `\#12\catcode `\^12\catcode `\_12\catcode `\%12\relax}%
\providecommand \@@startlink[1]{}%
\providecommand \@@endlink[0]{}%
\providecommand \url  [0]{\begingroup\@sanitize@url \@url }%
\providecommand \@url [1]{\endgroup\@href {#1}{\urlprefix }}%
\providecommand \urlprefix  [0]{URL }%
\providecommand \Eprint [0]{\href }%
\providecommand \doibase [0]{https://doi.org/}%
\providecommand \selectlanguage [0]{\@gobble}%
\providecommand \bibinfo  [0]{\@secondoftwo}%
\providecommand \bibfield  [0]{\@secondoftwo}%
\providecommand \translation [1]{[#1]}%
\providecommand \BibitemOpen [0]{}%
\providecommand \bibitemStop [0]{}%
\providecommand \bibitemNoStop [0]{.\EOS\space}%
\providecommand \EOS [0]{\spacefactor3000\relax}%
\providecommand \BibitemShut  [1]{\csname bibitem#1\endcsname}%
\let\auto@bib@innerbib\@empty
\bibitem [{\citenamefont {Abanin}\ \emph {et~al.}(2019)\citenamefont {Abanin},
  \citenamefont {Altman}, \citenamefont {Bloch},\ and\ \citenamefont
  {Serbyn}}]{Abanin_2019}%
  \BibitemOpen
  \bibfield  {author} {\bibinfo {author} {\bibfnamefont {D.~A.}\ \bibnamefont
  {Abanin}}, \bibinfo {author} {\bibfnamefont {E.}~\bibnamefont {Altman}},
  \bibinfo {author} {\bibfnamefont {I.}~\bibnamefont {Bloch}},\ and\ \bibinfo
  {author} {\bibfnamefont {M.}~\bibnamefont {Serbyn}},\ }\bibfield  {journal}
  {\bibinfo  {journal} {Reviews of Modern Physics}\ }\textbf {\bibinfo {volume}
  {91}},\ \href {https://doi.org/10.1103/revmodphys.91.021001}
  {10.1103/revmodphys.91.021001} (\bibinfo {year} {2019})\BibitemShut {NoStop}%
\bibitem [{\citenamefont {Swingle}\ \emph {et~al.}(2016)\citenamefont
  {Swingle}, \citenamefont {Bentsen}, \citenamefont {Schleier-Smith},\ and\
  \citenamefont {Hayden}}]{Swingle_2016}%
  \BibitemOpen
  \bibfield  {author} {\bibinfo {author} {\bibfnamefont {B.}~\bibnamefont
  {Swingle}}, \bibinfo {author} {\bibfnamefont {G.}~\bibnamefont {Bentsen}},
  \bibinfo {author} {\bibfnamefont {M.}~\bibnamefont {Schleier-Smith}},\ and\
  \bibinfo {author} {\bibfnamefont {P.}~\bibnamefont {Hayden}},\ }\bibfield
  {journal} {\bibinfo  {journal} {Physical Review A}\ }\textbf {\bibinfo
  {volume} {94}},\ \href {https://doi.org/10.1103/physreva.94.040302}
  {10.1103/physreva.94.040302} (\bibinfo {year} {2016})\BibitemShut {NoStop}%
\bibitem [{\citenamefont {Nandkishore}\ and\ \citenamefont
  {Huse}(2015)}]{Nandkishore_2015}%
  \BibitemOpen
  \bibfield  {author} {\bibinfo {author} {\bibfnamefont {R.}~\bibnamefont
  {Nandkishore}}\ and\ \bibinfo {author} {\bibfnamefont {D.~A.}\ \bibnamefont
  {Huse}},\ }\href {https://doi.org/10.1146/annurev-conmatphys-031214-014726}
  {\bibfield  {journal} {\bibinfo  {journal} {Annual Review of Condensed Matter
  Physics}\ }\textbf {\bibinfo {volume} {6}},\ \bibinfo {pages} {15–38}
  (\bibinfo {year} {2015})}\BibitemShut {NoStop}%
\bibitem [{\citenamefont {Khemani}\ \emph {et~al.}(2016)\citenamefont
  {Khemani}, \citenamefont {Lazarides}, \citenamefont {Moessner},\ and\
  \citenamefont {Sondhi}}]{Khemani_2016}%
  \BibitemOpen
  \bibfield  {author} {\bibinfo {author} {\bibfnamefont {V.}~\bibnamefont
  {Khemani}}, \bibinfo {author} {\bibfnamefont {A.}~\bibnamefont {Lazarides}},
  \bibinfo {author} {\bibfnamefont {R.}~\bibnamefont {Moessner}},\ and\
  \bibinfo {author} {\bibfnamefont {S.}~\bibnamefont {Sondhi}},\ }\bibfield
  {journal} {\bibinfo  {journal} {Physical Review Letters}\ }\textbf {\bibinfo
  {volume} {116}},\ \href {https://doi.org/10.1103/physrevlett.116.250401}
  {10.1103/physrevlett.116.250401} (\bibinfo {year} {2016})\BibitemShut
  {NoStop}%
\bibitem [{\citenamefont {Alba}\ and\ \citenamefont
  {Calabrese}(2017)}]{Alba_2017}%
  \BibitemOpen
  \bibfield  {author} {\bibinfo {author} {\bibfnamefont {V.}~\bibnamefont
  {Alba}}\ and\ \bibinfo {author} {\bibfnamefont {P.}~\bibnamefont
  {Calabrese}},\ }\href {https://doi.org/10.1073/pnas.1703516114} {\bibfield
  {journal} {\bibinfo  {journal} {Proceedings of the National Academy of
  Sciences}\ }\textbf {\bibinfo {volume} {114}},\ \bibinfo {pages}
  {7947–7951} (\bibinfo {year} {2017})}\BibitemShut {NoStop}%
\bibitem [{\citenamefont {Choi}\ \emph {et~al.}(2020)\citenamefont {Choi},
  \citenamefont {Bao}, \citenamefont {Qi},\ and\ \citenamefont
  {Altman}}]{Choi_2020}%
  \BibitemOpen
  \bibfield  {author} {\bibinfo {author} {\bibfnamefont {S.}~\bibnamefont
  {Choi}}, \bibinfo {author} {\bibfnamefont {Y.}~\bibnamefont {Bao}}, \bibinfo
  {author} {\bibfnamefont {X.-L.}\ \bibnamefont {Qi}},\ and\ \bibinfo {author}
  {\bibfnamefont {E.}~\bibnamefont {Altman}},\ }\bibfield  {journal} {\bibinfo
  {journal} {Physical Review Letters}\ }\textbf {\bibinfo {volume} {125}},\
  \href {https://doi.org/10.1103/physrevlett.125.030505}
  {10.1103/physrevlett.125.030505} (\bibinfo {year} {2020})\BibitemShut
  {NoStop}%
\bibitem [{\citenamefont {Heyl}(2018)}]{Heyl_2018}%
  \BibitemOpen
  \bibfield  {author} {\bibinfo {author} {\bibfnamefont {M.}~\bibnamefont
  {Heyl}},\ }\href {https://doi.org/10.1088/1361-6633/aaaf9a} {\bibfield
  {journal} {\bibinfo  {journal} {Reports on Progress in Physics}\ }\textbf
  {\bibinfo {volume} {81}},\ \bibinfo {pages} {054001} (\bibinfo {year}
  {2018})}\BibitemShut {NoStop}%
\bibitem [{\citenamefont {Hasan}\ and\ \citenamefont
  {Kane}(2010)}]{hasan_colloquium:_2010}%
  \BibitemOpen
  \bibfield  {author} {\bibinfo {author} {\bibfnamefont {M.~Z.}\ \bibnamefont
  {Hasan}}\ and\ \bibinfo {author} {\bibfnamefont {C.~L.}\ \bibnamefont
  {Kane}},\ }\href {https://doi.org/10.1103/RevModPhys.82.3045} {\bibfield
  {journal} {\bibinfo  {journal} {Rev. Mod. Phys.}\ }\textbf {\bibinfo {volume}
  {82}},\ \bibinfo {pages} {3045} (\bibinfo {year} {2010})}\BibitemShut
  {NoStop}%
\bibitem [{\citenamefont {Qi}\ and\ \citenamefont {Zhang}(2011)}]{qi_rmp_2011}%
  \BibitemOpen
  \bibfield  {author} {\bibinfo {author} {\bibfnamefont {X.-L.}\ \bibnamefont
  {Qi}}\ and\ \bibinfo {author} {\bibfnamefont {S.-C.}\ \bibnamefont {Zhang}},\
  }\href {https://doi.org/10.1103/RevModPhys.83.1057} {\bibfield  {journal}
  {\bibinfo  {journal} {Rev. Mod. Phys.}\ }\textbf {\bibinfo {volume} {83}},\
  \bibinfo {pages} {1057} (\bibinfo {year} {2011})}\BibitemShut {NoStop}%
\bibitem [{\citenamefont {Kitaev}\ \emph {et~al.}(2009)\citenamefont {Kitaev},
  \citenamefont {Lebedev},\ and\ \citenamefont {Feigel’man}}]{Kitaev_2009}%
  \BibitemOpen
  \bibfield  {author} {\bibinfo {author} {\bibfnamefont {A.}~\bibnamefont
  {Kitaev}}, \bibinfo {author} {\bibfnamefont {V.}~\bibnamefont {Lebedev}},\
  and\ \bibinfo {author} {\bibfnamefont {M.}~\bibnamefont {Feigel’man}},\
  }\bibfield  {journal} {\bibinfo  {journal} {AIP Conference Proceedings}\
  }\href {https://doi.org/10.1063/1.3149495} {10.1063/1.3149495} (\bibinfo
  {year} {2009})\BibitemShut {NoStop}%
\bibitem [{\citenamefont {Wang}\ \emph {et~al.}(2014)\citenamefont {Wang},
  \citenamefont {Potter},\ and\ \citenamefont {Senthil}}]{Wang_2014}%
  \BibitemOpen
  \bibfield  {author} {\bibinfo {author} {\bibfnamefont {C.}~\bibnamefont
  {Wang}}, \bibinfo {author} {\bibfnamefont {A.~C.}\ \bibnamefont {Potter}},\
  and\ \bibinfo {author} {\bibfnamefont {T.}~\bibnamefont {Senthil}},\ }\href
  {https://doi.org/10.1126/science.1243326} {\bibfield  {journal} {\bibinfo
  {journal} {Science}\ }\textbf {\bibinfo {volume} {343}},\ \bibinfo {pages}
  {629–631} (\bibinfo {year} {2014})}\BibitemShut {NoStop}%
\bibitem [{\citenamefont {Isobe}\ and\ \citenamefont {Fu}(2015)}]{Isobe_2015}%
  \BibitemOpen
  \bibfield  {author} {\bibinfo {author} {\bibfnamefont {H.}~\bibnamefont
  {Isobe}}\ and\ \bibinfo {author} {\bibfnamefont {L.}~\bibnamefont {Fu}},\
  }\bibfield  {journal} {\bibinfo  {journal} {Physical Review B}\ }\textbf
  {\bibinfo {volume} {92}},\ \href {https://doi.org/10.1103/physrevb.92.081304}
  {10.1103/physrevb.92.081304} (\bibinfo {year} {2015})\BibitemShut {NoStop}%
\bibitem [{\citenamefont {Schnyder}\ \emph {et~al.}(2008)\citenamefont
  {Schnyder}, \citenamefont {Ryu}, \citenamefont {Furusaki},\ and\
  \citenamefont {Ludwig}}]{classif_3d}%
  \BibitemOpen
  \bibfield  {author} {\bibinfo {author} {\bibfnamefont {A.~P.}\ \bibnamefont
  {Schnyder}}, \bibinfo {author} {\bibfnamefont {S.}~\bibnamefont {Ryu}},
  \bibinfo {author} {\bibfnamefont {A.}~\bibnamefont {Furusaki}},\ and\
  \bibinfo {author} {\bibfnamefont {A.~W.~W.}\ \bibnamefont {Ludwig}},\ }\href
  {https://doi.org/10.1103/PhysRevB.78.195125} {\bibfield  {journal} {\bibinfo
  {journal} {Phys. Rev. B}\ }\textbf {\bibinfo {volume} {78}},\ \bibinfo
  {pages} {195125} (\bibinfo {year} {2008})}\BibitemShut {NoStop}%
\bibitem [{\citenamefont {McGinley}\ and\ \citenamefont
  {Cooper}(2019{\natexlab{a}})}]{mcginley_classif_2019}%
  \BibitemOpen
  \bibfield  {author} {\bibinfo {author} {\bibfnamefont {M.}~\bibnamefont
  {McGinley}}\ and\ \bibinfo {author} {\bibfnamefont {N.~R.}\ \bibnamefont
  {Cooper}},\ }\bibfield  {journal} {\bibinfo  {journal} {Physical Review B}\
  }\textbf {\bibinfo {volume} {99}},\ \href
  {https://doi.org/10.1103/physrevb.99.075148} {10.1103/physrevb.99.075148}
  (\bibinfo {year} {2019}{\natexlab{a}})\BibitemShut {NoStop}%
\bibitem [{\citenamefont {McGinley}\ and\ \citenamefont
  {Cooper}(2019{\natexlab{b}})}]{mcginley_interacting_spt_noneq_2019}%
  \BibitemOpen
  \bibfield  {author} {\bibinfo {author} {\bibfnamefont {M.}~\bibnamefont
  {McGinley}}\ and\ \bibinfo {author} {\bibfnamefont {N.~R.}\ \bibnamefont
  {Cooper}},\ }\href {https://doi.org/10.1103/PhysRevResearch.1.033204}
  {\bibfield  {journal} {\bibinfo  {journal} {Phys. Rev. Research}\ }\textbf
  {\bibinfo {volume} {1}},\ \bibinfo {pages} {033204} (\bibinfo {year}
  {2019}{\natexlab{b}})}\BibitemShut {NoStop}%
\bibitem [{\citenamefont {McGinley}\ and\ \citenamefont
  {Cooper}(2018)}]{McGinley_1d_2018}%
  \BibitemOpen
  \bibfield  {author} {\bibinfo {author} {\bibfnamefont {M.}~\bibnamefont
  {McGinley}}\ and\ \bibinfo {author} {\bibfnamefont {N.~R.}\ \bibnamefont
  {Cooper}},\ }\bibfield  {journal} {\bibinfo  {journal} {Physical Review
  Letters}\ }\textbf {\bibinfo {volume} {121}},\ \href
  {https://doi.org/10.1103/physrevlett.121.090401}
  {10.1103/physrevlett.121.090401} (\bibinfo {year} {2018})\BibitemShut
  {NoStop}%
\bibitem [{\citenamefont {Schüler}\ and\ \citenamefont
  {Werner}(2017)}]{Schuler_2017}%
  \BibitemOpen
  \bibfield  {author} {\bibinfo {author} {\bibfnamefont {M.}~\bibnamefont
  {Schüler}}\ and\ \bibinfo {author} {\bibfnamefont {P.}~\bibnamefont
  {Werner}},\ }\bibfield  {journal} {\bibinfo  {journal} {Physical Review B}\
  }\textbf {\bibinfo {volume} {96}},\ \href
  {https://doi.org/10.1103/physrevb.96.155122} {10.1103/physrevb.96.155122}
  (\bibinfo {year} {2017})\BibitemShut {NoStop}%
\bibitem [{\citenamefont {Sun}\ \emph {et~al.}(2018)\citenamefont {Sun},
  \citenamefont {Yi}, \citenamefont {Wang}, \citenamefont {Zhang},
  \citenamefont {Sanders}, \citenamefont {Xu}, \citenamefont {Wang},
  \citenamefont {Schmiedmayer}, \citenamefont {Deng}, \citenamefont {Liu},\
  and\ \citenamefont {et~al.}}]{Sun_2018}%
  \BibitemOpen
  \bibfield  {author} {\bibinfo {author} {\bibfnamefont {W.}~\bibnamefont
  {Sun}}, \bibinfo {author} {\bibfnamefont {C.-R.}\ \bibnamefont {Yi}},
  \bibinfo {author} {\bibfnamefont {B.-Z.}\ \bibnamefont {Wang}}, \bibinfo
  {author} {\bibfnamefont {W.-W.}\ \bibnamefont {Zhang}}, \bibinfo {author}
  {\bibfnamefont {B.~C.}\ \bibnamefont {Sanders}}, \bibinfo {author}
  {\bibfnamefont {X.-T.}\ \bibnamefont {Xu}}, \bibinfo {author} {\bibfnamefont
  {Z.-Y.}\ \bibnamefont {Wang}}, \bibinfo {author} {\bibfnamefont
  {J.}~\bibnamefont {Schmiedmayer}}, \bibinfo {author} {\bibfnamefont
  {Y.}~\bibnamefont {Deng}}, \bibinfo {author} {\bibfnamefont {X.-J.}\
  \bibnamefont {Liu}},\ and\ \bibinfo {author} {\bibnamefont {et~al.}},\
  }\bibfield  {journal} {\bibinfo  {journal} {Physical Review Letters}\
  }\textbf {\bibinfo {volume} {121}},\ \href
  {https://doi.org/10.1103/physrevlett.121.250403}
  {10.1103/physrevlett.121.250403} (\bibinfo {year} {2018})\BibitemShut
  {NoStop}%
\bibitem [{\citenamefont {Hu}\ and\ \citenamefont
  {Zhao}(2020)}]{PhysRevLett.124.160402}%
  \BibitemOpen
  \bibfield  {author} {\bibinfo {author} {\bibfnamefont {H.}~\bibnamefont
  {Hu}}\ and\ \bibinfo {author} {\bibfnamefont {E.}~\bibnamefont {Zhao}},\
  }\href {https://doi.org/10.1103/PhysRevLett.124.160402} {\bibfield  {journal}
  {\bibinfo  {journal} {Phys. Rev. Lett.}\ }\textbf {\bibinfo {volume} {124}},\
  \bibinfo {pages} {160402} (\bibinfo {year} {2020})}\BibitemShut {NoStop}%
\bibitem [{\citenamefont {Tarnowski}\ \emph {et~al.}(2019)\citenamefont
  {Tarnowski}, \citenamefont {Ünal}, \citenamefont {Fläschner}, \citenamefont
  {Rem}, \citenamefont {Eckardt}, \citenamefont {Sengstock},\ and\
  \citenamefont {Weitenberg}}]{Tarnowski_2019}%
  \BibitemOpen
  \bibfield  {author} {\bibinfo {author} {\bibfnamefont {M.}~\bibnamefont
  {Tarnowski}}, \bibinfo {author} {\bibfnamefont {F.~N.}\ \bibnamefont
  {Ünal}}, \bibinfo {author} {\bibfnamefont {N.}~\bibnamefont {Fläschner}},
  \bibinfo {author} {\bibfnamefont {B.~S.}\ \bibnamefont {Rem}}, \bibinfo
  {author} {\bibfnamefont {A.}~\bibnamefont {Eckardt}}, \bibinfo {author}
  {\bibfnamefont {K.}~\bibnamefont {Sengstock}},\ and\ \bibinfo {author}
  {\bibfnamefont {C.}~\bibnamefont {Weitenberg}},\ }\bibfield  {journal}
  {\bibinfo  {journal} {Nature Communications}\ }\textbf {\bibinfo {volume}
  {10}},\ \href {https://doi.org/10.1038/s41467-019-09668-y}
  {10.1038/s41467-019-09668-y} (\bibinfo {year} {2019})\BibitemShut {NoStop}%
\bibitem [{\citenamefont {Tsomokos}\ \emph {et~al.}(2009)\citenamefont
  {Tsomokos}, \citenamefont {Hamma}, \citenamefont {Zhang}, \citenamefont
  {Haas},\ and\ \citenamefont {Fazio}}]{Tsomokos_2009}%
  \BibitemOpen
  \bibfield  {author} {\bibinfo {author} {\bibfnamefont {D.~I.}\ \bibnamefont
  {Tsomokos}}, \bibinfo {author} {\bibfnamefont {A.}~\bibnamefont {Hamma}},
  \bibinfo {author} {\bibfnamefont {W.}~\bibnamefont {Zhang}}, \bibinfo
  {author} {\bibfnamefont {S.}~\bibnamefont {Haas}},\ and\ \bibinfo {author}
  {\bibfnamefont {R.}~\bibnamefont {Fazio}},\ }\bibfield  {journal} {\bibinfo
  {journal} {Physical Review A}\ }\textbf {\bibinfo {volume} {80}},\ \href
  {https://doi.org/10.1103/physreva.80.060302} {10.1103/physreva.80.060302}
  (\bibinfo {year} {2009})\BibitemShut {NoStop}%
\bibitem [{\citenamefont {Kells}\ \emph {et~al.}(2014)\citenamefont {Kells},
  \citenamefont {Sen}, \citenamefont {Slingerland},\ and\ \citenamefont
  {Vishveshwara}}]{Kells_2014}%
  \BibitemOpen
  \bibfield  {author} {\bibinfo {author} {\bibfnamefont {G.}~\bibnamefont
  {Kells}}, \bibinfo {author} {\bibfnamefont {D.}~\bibnamefont {Sen}}, \bibinfo
  {author} {\bibfnamefont {J.~K.}\ \bibnamefont {Slingerland}},\ and\ \bibinfo
  {author} {\bibfnamefont {S.}~\bibnamefont {Vishveshwara}},\ }\bibfield
  {journal} {\bibinfo  {journal} {Physical Review B}\ }\textbf {\bibinfo
  {volume} {89}},\ \href {https://doi.org/10.1103/physrevb.89.235130}
  {10.1103/physrevb.89.235130} (\bibinfo {year} {2014})\BibitemShut {NoStop}%
\bibitem [{\citenamefont {Hauke}\ \emph {et~al.}(2014)\citenamefont {Hauke},
  \citenamefont {Lewenstein},\ and\ \citenamefont
  {Eckardt}}]{PhysRevLett.113.045303}%
  \BibitemOpen
  \bibfield  {author} {\bibinfo {author} {\bibfnamefont {P.}~\bibnamefont
  {Hauke}}, \bibinfo {author} {\bibfnamefont {M.}~\bibnamefont {Lewenstein}},\
  and\ \bibinfo {author} {\bibfnamefont {A.}~\bibnamefont {Eckardt}},\ }\href
  {https://doi.org/10.1103/PhysRevLett.113.045303} {\bibfield  {journal}
  {\bibinfo  {journal} {Phys. Rev. Lett.}\ }\textbf {\bibinfo {volume} {113}},\
  \bibinfo {pages} {045303} (\bibinfo {year} {2014})}\BibitemShut {NoStop}%
\bibitem [{\citenamefont {Pastori}\ \emph {et~al.}(2020)\citenamefont
  {Pastori}, \citenamefont {Barbarino},\ and\ \citenamefont
  {Budich}}]{Pastori_2020}%
  \BibitemOpen
  \bibfield  {author} {\bibinfo {author} {\bibfnamefont {L.}~\bibnamefont
  {Pastori}}, \bibinfo {author} {\bibfnamefont {S.}~\bibnamefont {Barbarino}},\
  and\ \bibinfo {author} {\bibfnamefont {J.~C.}\ \bibnamefont {Budich}},\
  }\bibfield  {journal} {\bibinfo  {journal} {Physical Review Research}\
  }\textbf {\bibinfo {volume} {2}},\ \href
  {https://doi.org/10.1103/physrevresearch.2.033259}
  {10.1103/physrevresearch.2.033259} (\bibinfo {year} {2020})\BibitemShut
  {NoStop}%
\bibitem [{\citenamefont {Sch\"uler}\ \emph {et~al.}(2019)\citenamefont
  {Sch\"uler}, \citenamefont {Budich},\ and\ \citenamefont
  {Werner}}]{PhysRevB.100.041101}%
  \BibitemOpen
  \bibfield  {author} {\bibinfo {author} {\bibfnamefont {M.}~\bibnamefont
  {Sch\"uler}}, \bibinfo {author} {\bibfnamefont {J.~C.}\ \bibnamefont
  {Budich}},\ and\ \bibinfo {author} {\bibfnamefont {P.}~\bibnamefont
  {Werner}},\ }\href {https://doi.org/10.1103/PhysRevB.100.041101} {\bibfield
  {journal} {\bibinfo  {journal} {Phys. Rev. B}\ }\textbf {\bibinfo {volume}
  {100}},\ \bibinfo {pages} {041101} (\bibinfo {year} {2019})}\BibitemShut
  {NoStop}%
\bibitem [{\citenamefont {Gonz\'alez-Cuadra}\ \emph {et~al.}(2019)\citenamefont
  {Gonz\'alez-Cuadra}, \citenamefont {Dauphin}, \citenamefont {Grzybowski},
  \citenamefont {W\'ojcik}, \citenamefont {Lewenstein},\ and\ \citenamefont
  {Bermudez}}]{cuadra_prb_2019}%
  \BibitemOpen
  \bibfield  {author} {\bibinfo {author} {\bibfnamefont {D.}~\bibnamefont
  {Gonz\'alez-Cuadra}}, \bibinfo {author} {\bibfnamefont {A.}~\bibnamefont
  {Dauphin}}, \bibinfo {author} {\bibfnamefont {P.~R.}\ \bibnamefont
  {Grzybowski}}, \bibinfo {author} {\bibfnamefont {P.}~\bibnamefont
  {W\'ojcik}}, \bibinfo {author} {\bibfnamefont {M.}~\bibnamefont
  {Lewenstein}},\ and\ \bibinfo {author} {\bibfnamefont {A.}~\bibnamefont
  {Bermudez}},\ }\href {https://doi.org/10.1103/PhysRevB.99.045139} {\bibfield
  {journal} {\bibinfo  {journal} {Phys. Rev. B}\ }\textbf {\bibinfo {volume}
  {99}},\ \bibinfo {pages} {045139} (\bibinfo {year} {2019})}\BibitemShut
  {NoStop}%
\bibitem [{\citenamefont {González-Cuadra}\ \emph {et~al.}(2019)\citenamefont
  {González-Cuadra}, \citenamefont {Bermudez}, \citenamefont {Grzybowski},
  \citenamefont {Lewenstein},\ and\ \citenamefont
  {Dauphin}}]{cuadra_natcomm_2019}%
  \BibitemOpen
  \bibfield  {author} {\bibinfo {author} {\bibfnamefont {D.}~\bibnamefont
  {González-Cuadra}}, \bibinfo {author} {\bibfnamefont {A.}~\bibnamefont
  {Bermudez}}, \bibinfo {author} {\bibfnamefont {P.~R.}\ \bibnamefont
  {Grzybowski}}, \bibinfo {author} {\bibfnamefont {M.}~\bibnamefont
  {Lewenstein}},\ and\ \bibinfo {author} {\bibfnamefont {A.}~\bibnamefont
  {Dauphin}},\ }\bibfield  {journal} {\bibinfo  {journal} {Nature
  Communications}\ }\textbf {\bibinfo {volume} {10}},\ \href
  {https://doi.org/10.1038/s41467-019-10796-8} {10.1038/s41467-019-10796-8}
  (\bibinfo {year} {2019})\BibitemShut {NoStop}%
\bibitem [{\citenamefont {Yu}(2017)}]{phase_vortices}%
  \BibitemOpen
  \bibfield  {author} {\bibinfo {author} {\bibfnamefont {J.}~\bibnamefont
  {Yu}},\ }\href {https://doi.org/10.1103/PhysRevA.96.023601} {\bibfield
  {journal} {\bibinfo  {journal} {Phys. Rev. A}\ }\textbf {\bibinfo {volume}
  {96}},\ \bibinfo {pages} {023601} (\bibinfo {year} {2017})}\BibitemShut
  {NoStop}%
\bibitem [{\citenamefont {Aidelsburger}\ \emph {et~al.}(2011)\citenamefont
  {Aidelsburger}, \citenamefont {Atala}, \citenamefont {Nascimb\`{e}ne},
  \citenamefont {Trotzky}, \citenamefont {Chen},\ and\ \citenamefont
  {Bloch}}]{aidelsburger_experimental_2011}%
  \BibitemOpen
  \bibfield  {author} {\bibinfo {author} {\bibfnamefont {M.}~\bibnamefont
  {Aidelsburger}}, \bibinfo {author} {\bibfnamefont {M.}~\bibnamefont {Atala}},
  \bibinfo {author} {\bibfnamefont {S.}~\bibnamefont {Nascimb\`{e}ne}},
  \bibinfo {author} {\bibfnamefont {S.}~\bibnamefont {Trotzky}}, \bibinfo
  {author} {\bibfnamefont {Y.-A.}\ \bibnamefont {Chen}},\ and\ \bibinfo
  {author} {\bibfnamefont {I.}~\bibnamefont {Bloch}},\ }\href
  {https://doi.org/10.1103/PhysRevLett.107.255301} {\bibfield  {journal}
  {\bibinfo  {journal} {Phys. Rev. Lett.}\ }\textbf {\bibinfo {volume} {107}},\
  \bibinfo {pages} {255301} (\bibinfo {year} {2011})}\BibitemShut {NoStop}%
\bibitem [{\citenamefont {Struck}\ \emph {et~al.}(2012)\citenamefont {Struck},
  \citenamefont {\"{O}lschl\"{a}ger}, \citenamefont {Weinberg}, \citenamefont
  {Hauke}, \citenamefont {Simonet}, \citenamefont {Eckardt}, \citenamefont
  {Lewenstein}, \citenamefont {Sengstock},\ and\ \citenamefont
  {Windpassinger}}]{struck_tunable_2012}%
  \BibitemOpen
  \bibfield  {author} {\bibinfo {author} {\bibfnamefont {J.}~\bibnamefont
  {Struck}}, \bibinfo {author} {\bibfnamefont {C.}~\bibnamefont
  {\"{O}lschl\"{a}ger}}, \bibinfo {author} {\bibfnamefont {M.}~\bibnamefont
  {Weinberg}}, \bibinfo {author} {\bibfnamefont {P.}~\bibnamefont {Hauke}},
  \bibinfo {author} {\bibfnamefont {J.}~\bibnamefont {Simonet}}, \bibinfo
  {author} {\bibfnamefont {A.}~\bibnamefont {Eckardt}}, \bibinfo {author}
  {\bibfnamefont {M.}~\bibnamefont {Lewenstein}}, \bibinfo {author}
  {\bibfnamefont {K.}~\bibnamefont {Sengstock}},\ and\ \bibinfo {author}
  {\bibfnamefont {P.}~\bibnamefont {Windpassinger}},\ }\href
  {https://doi.org/10.1103/PhysRevLett.108.225304} {\bibfield  {journal}
  {\bibinfo  {journal} {Phys. Rev. Lett.}\ }\textbf {\bibinfo {volume} {108}},\
  \bibinfo {pages} {225304} (\bibinfo {year} {2012})}\BibitemShut {NoStop}%
\bibitem [{\citenamefont {Goldman}\ \emph {et~al.}(2016)\citenamefont
  {Goldman}, \citenamefont {Budich},\ and\ \citenamefont
  {Zoller}}]{goldman_topological_2016}%
  \BibitemOpen
  \bibfield  {author} {\bibinfo {author} {\bibfnamefont {N.}~\bibnamefont
  {Goldman}}, \bibinfo {author} {\bibfnamefont {J.~C.}\ \bibnamefont
  {Budich}},\ and\ \bibinfo {author} {\bibfnamefont {P.}~\bibnamefont
  {Zoller}},\ }\href {https://doi.org/10.1038/nphys3803} {\bibfield  {journal}
  {\bibinfo  {journal} {Nature Phys.}\ }\textbf {\bibinfo {volume} {12}},\
  \bibinfo {pages} {639} (\bibinfo {year} {2016})}\BibitemShut {NoStop}%
\bibitem [{\citenamefont {Dauphin}\ \emph {et~al.}(2017)\citenamefont
  {Dauphin}, \citenamefont {Tran}, \citenamefont {Lewenstein},\ and\
  \citenamefont {Goldman}}]{dauphin_loading_2017}%
  \BibitemOpen
  \bibfield  {author} {\bibinfo {author} {\bibfnamefont {A.}~\bibnamefont
  {Dauphin}}, \bibinfo {author} {\bibfnamefont {D.-T.}\ \bibnamefont {Tran}},
  \bibinfo {author} {\bibfnamefont {M.}~\bibnamefont {Lewenstein}},\ and\
  \bibinfo {author} {\bibfnamefont {N.}~\bibnamefont {Goldman}},\ }\href
  {https://doi.org/10.1088/2053-1583/aa6a3b} {\bibfield  {journal} {\bibinfo
  {journal} {2D Mater.}\ }\textbf {\bibinfo {volume} {4}},\ \bibinfo {pages}
  {024010} (\bibinfo {year} {2017})}\BibitemShut {NoStop}%
\bibitem [{\citenamefont {Fl\"{a}schner}\ \emph {et~al.}(2016)\citenamefont
  {Fl\"{a}schner}, \citenamefont {Rem}, \citenamefont {Tarnowski},
  \citenamefont {Vogel}, \citenamefont {L\"{u}hmann}, \citenamefont
  {Sengstock},\ and\ \citenamefont {Weitenberg}}]{flaschner_experimental_2016}%
  \BibitemOpen
  \bibfield  {author} {\bibinfo {author} {\bibfnamefont {N.}~\bibnamefont
  {Fl\"{a}schner}}, \bibinfo {author} {\bibfnamefont {B.~S.}\ \bibnamefont
  {Rem}}, \bibinfo {author} {\bibfnamefont {M.}~\bibnamefont {Tarnowski}},
  \bibinfo {author} {\bibfnamefont {D.}~\bibnamefont {Vogel}}, \bibinfo
  {author} {\bibfnamefont {D.-S.}\ \bibnamefont {L\"{u}hmann}}, \bibinfo
  {author} {\bibfnamefont {K.}~\bibnamefont {Sengstock}},\ and\ \bibinfo
  {author} {\bibfnamefont {C.}~\bibnamefont {Weitenberg}},\ }\href
  {https://doi.org/10.1126/science.aad4568} {\bibfield  {journal} {\bibinfo
  {journal} {Science}\ }\textbf {\bibinfo {volume} {352}},\ \bibinfo {pages}
  {1091} (\bibinfo {year} {2016})}\BibitemShut {NoStop}%
\bibitem [{\citenamefont {Atala}\ \emph {et~al.}(2013)\citenamefont {Atala},
  \citenamefont {Aidelsburger}, \citenamefont {Barreiro}, \citenamefont
  {Abanin}, \citenamefont {Kitagawa}, \citenamefont {Demler},\ and\
  \citenamefont {Bloch}}]{Atala_2013}%
  \BibitemOpen
  \bibfield  {author} {\bibinfo {author} {\bibfnamefont {M.}~\bibnamefont
  {Atala}}, \bibinfo {author} {\bibfnamefont {M.}~\bibnamefont {Aidelsburger}},
  \bibinfo {author} {\bibfnamefont {J.~T.}\ \bibnamefont {Barreiro}}, \bibinfo
  {author} {\bibfnamefont {D.}~\bibnamefont {Abanin}}, \bibinfo {author}
  {\bibfnamefont {T.}~\bibnamefont {Kitagawa}}, \bibinfo {author}
  {\bibfnamefont {E.}~\bibnamefont {Demler}},\ and\ \bibinfo {author}
  {\bibfnamefont {I.}~\bibnamefont {Bloch}},\ }\href
  {https://doi.org/10.1038/nphys2790} {\bibfield  {journal} {\bibinfo
  {journal} {Nature Physics}\ }\textbf {\bibinfo {volume} {9}},\ \bibinfo
  {pages} {795–800} (\bibinfo {year} {2013})}\BibitemShut {NoStop}%
\bibitem [{\citenamefont {Wang}\ \emph {et~al.}(2016)\citenamefont {Wang},
  \citenamefont {Xiao}, \citenamefont {Liu}, \citenamefont {Zhu},\ and\
  \citenamefont {Chan}}]{Wang_2016}%
  \BibitemOpen
  \bibfield  {author} {\bibinfo {author} {\bibfnamefont {Q.}~\bibnamefont
  {Wang}}, \bibinfo {author} {\bibfnamefont {M.}~\bibnamefont {Xiao}}, \bibinfo
  {author} {\bibfnamefont {H.}~\bibnamefont {Liu}}, \bibinfo {author}
  {\bibfnamefont {S.}~\bibnamefont {Zhu}},\ and\ \bibinfo {author}
  {\bibfnamefont {C.~T.}\ \bibnamefont {Chan}},\ }\bibfield  {journal}
  {\bibinfo  {journal} {Physical Review B}\ }\textbf {\bibinfo {volume} {93}},\
  \href {https://doi.org/10.1103/physrevb.93.041415}
  {10.1103/physrevb.93.041415} (\bibinfo {year} {2016})\BibitemShut {NoStop}%
\bibitem [{\citenamefont {Zheng}\ \emph {et~al.}(2020)\citenamefont {Zheng},
  \citenamefont {Irsigler}, \citenamefont {Jiang}, \citenamefont {Weitenberg},\
  and\ \citenamefont {Hofstetter}}]{Zheng_2020}%
  \BibitemOpen
  \bibfield  {author} {\bibinfo {author} {\bibfnamefont {J.-H.}\ \bibnamefont
  {Zheng}}, \bibinfo {author} {\bibfnamefont {B.}~\bibnamefont {Irsigler}},
  \bibinfo {author} {\bibfnamefont {L.}~\bibnamefont {Jiang}}, \bibinfo
  {author} {\bibfnamefont {C.}~\bibnamefont {Weitenberg}},\ and\ \bibinfo
  {author} {\bibfnamefont {W.}~\bibnamefont {Hofstetter}},\ }\bibfield
  {journal} {\bibinfo  {journal} {Physical Review A}\ }\textbf {\bibinfo
  {volume} {101}},\ \href {https://doi.org/10.1103/physreva.101.013631}
  {10.1103/physreva.101.013631} (\bibinfo {year} {2020})\BibitemShut {NoStop}%
\bibitem [{\citenamefont {Peña~Ardila}\ \emph {et~al.}(2018)\citenamefont
  {Peña~Ardila}, \citenamefont {Heyl},\ and\ \citenamefont
  {Eckardt}}]{Pe_a_Ardila_2018}%
  \BibitemOpen
  \bibfield  {author} {\bibinfo {author} {\bibfnamefont {L.~A.}\ \bibnamefont
  {Peña~Ardila}}, \bibinfo {author} {\bibfnamefont {M.}~\bibnamefont {Heyl}},\
  and\ \bibinfo {author} {\bibfnamefont {A.}~\bibnamefont {Eckardt}},\
  }\bibfield  {journal} {\bibinfo  {journal} {Physical Review Letters}\
  }\textbf {\bibinfo {volume} {121}},\ \href
  {https://doi.org/10.1103/physrevlett.121.260401}
  {10.1103/physrevlett.121.260401} (\bibinfo {year} {2018})\BibitemShut
  {NoStop}%
\bibitem [{\citenamefont {Aidelsburger}\ \emph {et~al.}(2014)\citenamefont
  {Aidelsburger}, \citenamefont {Lohse}, \citenamefont {Schweizer},
  \citenamefont {Atala}, \citenamefont {Barreiro}, \citenamefont {Nascimbène},
  \citenamefont {Cooper}, \citenamefont {Bloch},\ and\ \citenamefont
  {Goldman}}]{Aidelsburger_2014}%
  \BibitemOpen
  \bibfield  {author} {\bibinfo {author} {\bibfnamefont {M.}~\bibnamefont
  {Aidelsburger}}, \bibinfo {author} {\bibfnamefont {M.}~\bibnamefont {Lohse}},
  \bibinfo {author} {\bibfnamefont {C.}~\bibnamefont {Schweizer}}, \bibinfo
  {author} {\bibfnamefont {M.}~\bibnamefont {Atala}}, \bibinfo {author}
  {\bibfnamefont {J.}~\bibnamefont {Barreiro}}, \bibinfo {author}
  {\bibfnamefont {S.}~\bibnamefont {Nascimbène}}, \bibinfo {author}
  {\bibfnamefont {N.}~\bibnamefont {Cooper}}, \bibinfo {author} {\bibfnamefont
  {I.}~\bibnamefont {Bloch}},\ and\ \bibinfo {author} {\bibfnamefont
  {N.}~\bibnamefont {Goldman}},\ }\href {https://doi.org/10.1038/nphys3171}
  {\bibfield  {journal} {\bibinfo  {journal} {Nature Physics}\ }\textbf
  {\bibinfo {volume} {11}},\ \bibinfo {pages} {162–166} (\bibinfo {year}
  {2014})}\BibitemShut {NoStop}%
\bibitem [{\citenamefont {Elben}\ \emph {et~al.}(2020)\citenamefont {Elben},
  \citenamefont {Yu}, \citenamefont {Zhu}, \citenamefont {Hafezi},
  \citenamefont {Pollmann}, \citenamefont {Zoller},\ and\ \citenamefont
  {Vermersch}}]{Elben_2020}%
  \BibitemOpen
  \bibfield  {author} {\bibinfo {author} {\bibfnamefont {A.}~\bibnamefont
  {Elben}}, \bibinfo {author} {\bibfnamefont {J.}~\bibnamefont {Yu}}, \bibinfo
  {author} {\bibfnamefont {G.}~\bibnamefont {Zhu}}, \bibinfo {author}
  {\bibfnamefont {M.}~\bibnamefont {Hafezi}}, \bibinfo {author} {\bibfnamefont
  {F.}~\bibnamefont {Pollmann}}, \bibinfo {author} {\bibfnamefont
  {P.}~\bibnamefont {Zoller}},\ and\ \bibinfo {author} {\bibfnamefont
  {B.}~\bibnamefont {Vermersch}},\ }\href
  {https://doi.org/10.1126/sciadv.aaz3666} {\bibfield  {journal} {\bibinfo
  {journal} {Science Advances}\ }\textbf {\bibinfo {volume} {6}},\ \bibinfo
  {pages} {eaaz3666} (\bibinfo {year} {2020})}\BibitemShut {NoStop}%
\bibitem [{\citenamefont {Dehghani}\ \emph {et~al.}(2020)\citenamefont
  {Dehghani}, \citenamefont {Cian}, \citenamefont {Hafezi},\ and\ \citenamefont
  {Barkeshli}}]{dehghani2020extraction}%
  \BibitemOpen
  \bibfield  {author} {\bibinfo {author} {\bibfnamefont {H.}~\bibnamefont
  {Dehghani}}, \bibinfo {author} {\bibfnamefont {Z.-P.}\ \bibnamefont {Cian}},
  \bibinfo {author} {\bibfnamefont {M.}~\bibnamefont {Hafezi}},\ and\ \bibinfo
  {author} {\bibfnamefont {M.}~\bibnamefont {Barkeshli}},\ }\href@noop {}
  {\bibinfo {title} {Extraction of many-body chern number from a single wave
  function}} (\bibinfo {year} {2020}),\ \Eprint
  {https://arxiv.org/abs/2005.13677} {arXiv:2005.13677 [cond-mat.str-el]}
  \BibitemShut {NoStop}%
\bibitem [{\citenamefont {Budich}\ and\ \citenamefont
  {Heyl}(2016)}]{PhysRevB.93.085416}%
  \BibitemOpen
  \bibfield  {author} {\bibinfo {author} {\bibfnamefont {J.~C.}\ \bibnamefont
  {Budich}}\ and\ \bibinfo {author} {\bibfnamefont {M.}~\bibnamefont {Heyl}},\
  }\href {https://doi.org/10.1103/PhysRevB.93.085416} {\bibfield  {journal}
  {\bibinfo  {journal} {Phys. Rev. B}\ }\textbf {\bibinfo {volume} {93}},\
  \bibinfo {pages} {085416} (\bibinfo {year} {2016})}\BibitemShut {NoStop}%
\bibitem [{\citenamefont {Kruckenhauser}\ and\ \citenamefont
  {Budich}(2018)}]{kruckenhauser_dynamical_2018}%
  \BibitemOpen
  \bibfield  {author} {\bibinfo {author} {\bibfnamefont {A.}~\bibnamefont
  {Kruckenhauser}}\ and\ \bibinfo {author} {\bibfnamefont {J.~C.}\ \bibnamefont
  {Budich}},\ }\href {https://doi.org/10.1103/PhysRevB.98.195124} {\bibfield
  {journal} {\bibinfo  {journal} {Phys. Rev. B}\ }\textbf {\bibinfo {volume}
  {98}},\ \bibinfo {pages} {195124} (\bibinfo {year} {2018})}\BibitemShut
  {NoStop}%
\bibitem [{\citenamefont {Su}\ \emph {et~al.}(1979)\citenamefont {Su},
  \citenamefont {Schrieffer},\ and\ \citenamefont {Heeger}}]{su_solitons_1979}%
  \BibitemOpen
  \bibfield  {author} {\bibinfo {author} {\bibfnamefont {W.~P.}\ \bibnamefont
  {Su}}, \bibinfo {author} {\bibfnamefont {J.~R.}\ \bibnamefont {Schrieffer}},\
  and\ \bibinfo {author} {\bibfnamefont {A.~J.}\ \bibnamefont {Heeger}},\
  }\href {https://doi.org/10.1103/PhysRevLett.42.1698} {\bibfield  {journal}
  {\bibinfo  {journal} {Phys. Rev. Lett.}\ }\textbf {\bibinfo {volume} {42}},\
  \bibinfo {pages} {1698} (\bibinfo {year} {1979})}\BibitemShut {NoStop}%
\bibitem [{\citenamefont {Heeger}\ \emph {et~al.}(1988)\citenamefont {Heeger},
  \citenamefont {Kivelson}, \citenamefont {Schrieffer},\ and\ \citenamefont
  {Su}}]{heeger_solitons_1988}%
  \BibitemOpen
  \bibfield  {author} {\bibinfo {author} {\bibfnamefont {A.~J.}\ \bibnamefont
  {Heeger}}, \bibinfo {author} {\bibfnamefont {S.}~\bibnamefont {Kivelson}},
  \bibinfo {author} {\bibfnamefont {J.~R.}\ \bibnamefont {Schrieffer}},\ and\
  \bibinfo {author} {\bibfnamefont {W.~P.}\ \bibnamefont {Su}},\ }\href
  {https://doi.org/10.1103/RevModPhys.60.781} {\bibfield  {journal} {\bibinfo
  {journal} {Rev. Mod. Phys.}\ }\textbf {\bibinfo {volume} {60}},\ \bibinfo
  {pages} {781} (\bibinfo {year} {1988})}\BibitemShut {NoStop}%
\bibitem [{\citenamefont {Marks}\ \emph {et~al.}(2021)\citenamefont {Marks},
  \citenamefont {Sch\"uler}, \citenamefont {Budich},\ and\ \citenamefont
  {Devereaux}}]{marks2020correlationassisted}%
  \BibitemOpen
  \bibfield  {author} {\bibinfo {author} {\bibfnamefont {J.~A.}\ \bibnamefont
  {Marks}}, \bibinfo {author} {\bibfnamefont {M.}~\bibnamefont {Sch\"uler}},
  \bibinfo {author} {\bibfnamefont {J.~C.}\ \bibnamefont {Budich}},\ and\
  \bibinfo {author} {\bibfnamefont {T.~P.}\ \bibnamefont {Devereaux}},\ }\href
  {https://doi.org/10.1103/PhysRevB.103.035112} {\bibfield  {journal} {\bibinfo
   {journal} {Phys. Rev. B}\ }\textbf {\bibinfo {volume} {103}},\ \bibinfo
  {pages} {035112} (\bibinfo {year} {2021})}\BibitemShut {NoStop}%
\bibitem [{Note1()}]{Note1}%
  \BibitemOpen
  \bibinfo {note} {In particular, quantized charge pumping reveals the
  underlying topology via adiabatic cyclic evolution. In fact, the obstructing
  order can even \protect \textit {drive} charge pumping.}\BibitemShut {Stop}%
\bibitem [{Note2()}]{Note2}%
  \BibitemOpen
  \bibinfo {note} {According to the Mermin-Wagner theorem, this is the case for
  all one-dimensional quantum systems}\BibitemShut {NoStop}%
\bibitem [{\citenamefont {Thouless}\ \emph {et~al.}(1982)\citenamefont
  {Thouless}, \citenamefont {Kohmoto}, \citenamefont {Nightingale},\ and\
  \citenamefont {den Nijs}}]{tknn_1982}%
  \BibitemOpen
  \bibfield  {author} {\bibinfo {author} {\bibfnamefont {D.~J.}\ \bibnamefont
  {Thouless}}, \bibinfo {author} {\bibfnamefont {M.}~\bibnamefont {Kohmoto}},
  \bibinfo {author} {\bibfnamefont {M.~P.}\ \bibnamefont {Nightingale}},\ and\
  \bibinfo {author} {\bibfnamefont {M.}~\bibnamefont {den Nijs}},\ }\href
  {https://doi.org/10.1103/PhysRevLett.49.405} {\bibfield  {journal} {\bibinfo
  {journal} {Phys. Rev. Lett.}\ }\textbf {\bibinfo {volume} {49}},\ \bibinfo
  {pages} {405} (\bibinfo {year} {1982})}\BibitemShut {NoStop}%
\bibitem [{\citenamefont {Avron}\ \emph {et~al.}(1983)\citenamefont {Avron},
  \citenamefont {Seiler},\ and\ \citenamefont {Simon}}]{PhysRevLett.51.51}%
  \BibitemOpen
  \bibfield  {author} {\bibinfo {author} {\bibfnamefont {J.~E.}\ \bibnamefont
  {Avron}}, \bibinfo {author} {\bibfnamefont {R.}~\bibnamefont {Seiler}},\ and\
  \bibinfo {author} {\bibfnamefont {B.}~\bibnamefont {Simon}},\ }\href
  {https://doi.org/10.1103/PhysRevLett.51.51} {\bibfield  {journal} {\bibinfo
  {journal} {Phys. Rev. Lett.}\ }\textbf {\bibinfo {volume} {51}},\ \bibinfo
  {pages} {51} (\bibinfo {year} {1983})}\BibitemShut {NoStop}%
\bibitem [{\citenamefont {Shapourian}\ \emph {et~al.}(2017)\citenamefont
  {Shapourian}, \citenamefont {Shiozaki},\ and\ \citenamefont
  {Ryu}}]{Shapourian_2017}%
  \BibitemOpen
  \bibfield  {author} {\bibinfo {author} {\bibfnamefont {H.}~\bibnamefont
  {Shapourian}}, \bibinfo {author} {\bibfnamefont {K.}~\bibnamefont
  {Shiozaki}},\ and\ \bibinfo {author} {\bibfnamefont {S.}~\bibnamefont
  {Ryu}},\ }\bibfield  {journal} {\bibinfo  {journal} {Physical Review
  Letters}\ }\textbf {\bibinfo {volume} {118}},\ \href
  {https://doi.org/10.1103/physrevlett.118.216402}
  {10.1103/physrevlett.118.216402} (\bibinfo {year} {2017})\BibitemShut
  {NoStop}%
\bibitem [{\citenamefont {Hung}\ and\ \citenamefont {Wen}(2014)}]{Hung_2014}%
  \BibitemOpen
  \bibfield  {author} {\bibinfo {author} {\bibfnamefont {L.-Y.}\ \bibnamefont
  {Hung}}\ and\ \bibinfo {author} {\bibfnamefont {X.-G.}\ \bibnamefont {Wen}},\
  }\bibfield  {journal} {\bibinfo  {journal} {Physical Review B}\ }\textbf
  {\bibinfo {volume} {89}},\ \href {https://doi.org/10.1103/physrevb.89.075121}
  {10.1103/physrevb.89.075121} (\bibinfo {year} {2014})\BibitemShut {NoStop}%
\bibitem [{\citenamefont {Zak}(1989)}]{zak_berrys_1989}%
  \BibitemOpen
  \bibfield  {author} {\bibinfo {author} {\bibfnamefont {J.}~\bibnamefont
  {Zak}},\ }\href {https://doi.org/10.1103/PhysRevLett.62.2747} {\bibfield
  {journal} {\bibinfo  {journal} {Phys. Rev. Lett.}\ }\textbf {\bibinfo
  {volume} {62}},\ \bibinfo {pages} {2747} (\bibinfo {year}
  {1989})}\BibitemShut {NoStop}%
\bibitem [{\citenamefont
  {Resta}(1998{\natexlab{a}})}]{resta_quantum-mechanical_1998}%
  \BibitemOpen
  \bibfield  {author} {\bibinfo {author} {\bibfnamefont {R.}~\bibnamefont
  {Resta}},\ }\href {https://doi.org/10.1103/PhysRevLett.80.1800} {\bibfield
  {journal} {\bibinfo  {journal} {Phys. Rev. Lett.}\ }\textbf {\bibinfo
  {volume} {80}},\ \bibinfo {pages} {1800} (\bibinfo {year}
  {1998}{\natexlab{a}})}\BibitemShut {NoStop}%
\bibitem [{\citenamefont {Ortiz}\ and\ \citenamefont
  {Martin}(1994)}]{pol_berry_phase}%
  \BibitemOpen
  \bibfield  {author} {\bibinfo {author} {\bibfnamefont {G.}~\bibnamefont
  {Ortiz}}\ and\ \bibinfo {author} {\bibfnamefont {R.~M.}\ \bibnamefont
  {Martin}},\ }\href {https://doi.org/10.1103/PhysRevB.49.14202} {\bibfield
  {journal} {\bibinfo  {journal} {Phys. Rev. B}\ }\textbf {\bibinfo {volume}
  {49}},\ \bibinfo {pages} {14202} (\bibinfo {year} {1994})}\BibitemShut
  {NoStop}%
\bibitem [{\citenamefont
  {Resta}(1998{\natexlab{b}})}]{Resta_1998_qm_pos_extended}%
  \BibitemOpen
  \bibfield  {author} {\bibinfo {author} {\bibfnamefont {R.}~\bibnamefont
  {Resta}},\ }\href {https://doi.org/10.1103/physrevlett.80.1800} {\bibfield
  {journal} {\bibinfo  {journal} {Physical Review Letters}\ }\textbf {\bibinfo
  {volume} {80}},\ \bibinfo {pages} {1800–1803} (\bibinfo {year}
  {1998}{\natexlab{b}})}\BibitemShut {NoStop}%
\bibitem [{Note3()}]{Note3}%
  \BibitemOpen
  \bibinfo {note} {We assume tacitly that $\protect \hat {H}(J, d, \tau , V,
  \Delta =0)$ is not order-obstructed.}\BibitemShut {Stop}%
\bibitem [{Note4()}]{Note4}%
  \BibitemOpen
  \bibinfo {note} {We note that the ratio $t_2/t_1$ is dimensionless in its own
  right, and thus invariant to multiplicative factors applied to the entire
  quench Hamiltonian. One could just as well have worked with the bare ratio.
  We find its logarithm to align better with intuitive notions of
  stability.}\BibitemShut {Stop}%
\bibitem [{\citenamefont {Weinberg}\ and\ \citenamefont
  {Bukov}(2017)}]{quspin1}%
  \BibitemOpen
  \bibfield  {author} {\bibinfo {author} {\bibfnamefont {P.}~\bibnamefont
  {Weinberg}}\ and\ \bibinfo {author} {\bibfnamefont {M.}~\bibnamefont
  {Bukov}},\ }\href {https://doi.org/10.21468/SciPostPhys.2.1.003} {\bibfield
  {journal} {\bibinfo  {journal} {SciPost Phys.}\ }\textbf {\bibinfo {volume}
  {2}},\ \bibinfo {pages} {003} (\bibinfo {year} {2017})}\BibitemShut {NoStop}%
\bibitem [{\citenamefont {Weinberg}\ and\ \citenamefont
  {Bukov}(2019)}]{quspin2}%
  \BibitemOpen
  \bibfield  {author} {\bibinfo {author} {\bibfnamefont {P.}~\bibnamefont
  {Weinberg}}\ and\ \bibinfo {author} {\bibfnamefont {M.}~\bibnamefont
  {Bukov}},\ }\href {https://doi.org/10.21468/SciPostPhys.7.2.020} {\bibfield
  {journal} {\bibinfo  {journal} {SciPost Phys.}\ }\textbf {\bibinfo {volume}
  {7}},\ \bibinfo {pages} {20} (\bibinfo {year} {2019})}\BibitemShut {NoStop}%
\bibitem [{\citenamefont {Coh}\ and\ \citenamefont {Vanderbilt}()}]{pythtb}%
  \BibitemOpen
  \bibfield  {author} {\bibinfo {author} {\bibfnamefont {S.}~\bibnamefont
  {Coh}}\ and\ \bibinfo {author} {\bibfnamefont {D.}~\bibnamefont
  {Vanderbilt}},\ }\href@noop {} {\bibinfo {title} {{P}ython {T}ight {B}inding
  {P}ackage}},\ \bibinfo {howpublished}
  {\url{physics.rutgers.edu/pythtb}}\BibitemShut {NoStop}%
\bibitem [{\citenamefont {Fishman}\ \emph {et~al.}(2020)\citenamefont
  {Fishman}, \citenamefont {White},\ and\ \citenamefont
  {Stoudenmire}}]{itensor}%
  \BibitemOpen
  \bibfield  {author} {\bibinfo {author} {\bibfnamefont {M.}~\bibnamefont
  {Fishman}}, \bibinfo {author} {\bibfnamefont {S.~R.}\ \bibnamefont {White}},\
  and\ \bibinfo {author} {\bibfnamefont {E.~M.}\ \bibnamefont {Stoudenmire}},\
  }\href@noop {} {\bibinfo {title} {The {ITENSOR} {S}oftware {L}ibrary for
  {T}ensor {N}etwork {C}alculations}} (\bibinfo {year} {2020}),\ \Eprint
  {https://arxiv.org/abs/2007.14822} {arXiv:2007.14822} \BibitemShut {NoStop}%
\end{thebibliography}

%




\end{document}